\documentclass{stars-techreport}

\usepackage{titlesec}
\usepackage{graphicx}
\graphicspath{{figs/}}
\usepackage{xcolor}

\usepackage{amsmath}
\usepackage{amssymb}
\usepackage{amsthm}
\usepackage{mathrsfs}
\usepackage{mathtools}
\usepackage{bm}
\allowdisplaybreaks

\let\originalleft\left
\let\originalright\right
\renewcommand{\left}{\mathopen{}\mathclose\bgroup\originalleft}
\renewcommand{\right}{\aftergroup\egroup\originalright}

\theoremstyle{definition}
\newtheorem{identity}{Identity}[section]

\usepackage{float}
\floatstyle{plain}
\newfloat{infobox}{thb}{none}

\usepackage[hang,flushmargin]{footmisc}

\usepackage{mdframed}
\mdfsetup{%
  backgroundcolor=gray!12,
  innerleftmargin=10pt,
  innertopmargin=10pt,
  innerrightmargin=10pt,
  innerbottommargin=10pt,
  skipabove=10pt
}

\usepackage[backend=biber,hyperref=true,style=ieee-caps]{biblatex}
\usepackage[noabbrev]{cleveref}
\crefformat{equation}{(#2#1#3)}
\Crefformat{equation}{(#2#1#3)}
\crefrangeformat{equation}{(#3#1#4) to~(#5#2#6)}
\Crefrangeformat{equation}{(#3#1#4) to~(#5#2#6)}
\crefmultiformat{equation}{(#2#1#3)}{ and~(#2#1#3)}{, (#2#1#3)}{, and~(#2#1#3)}
\Crefmultiformat{equation}{(#2#1#3)}{ and~(#2#1#3)}{, (#2#1#3)}{, and~(#2#1#3)}
\crefname{identity}{Identity}{Identities}

\AddToHook{cmd/appendix/before}{\crefalias{section}{appendix}}

\usepackage{xparse}

\NewDocumentCommand\bbm{}{ \begin{bmatrix} }
\NewDocumentCommand\ebm{}{ \end{bmatrix} }
\NewDocumentCommand\Vector{m}{ \boldsymbol{\mathbf{#1}} }

\NewDocumentCommand\Matrix{m}{ \bm{\mathbf{#1}} }
\NewDocumentCommand\Transpose{m}{ \left.{#1}\right.^\T }

\NewDocumentCommand\Trace{m}{ \mathrm{tr}\left(#1\right) }
\NewDocumentCommand\Determinant{m}{ \mathrm{det}\left(#1\right) }
\NewDocumentCommand\Norm{m}{ \left\Vert#1\right\Vert }

\NewDocumentCommand\at{mm}{\left.#1\right|_{#2}}

\NewDocumentCommand\Real{}{ \mathbb{R} }

\NewDocumentCommand\LieGroupAny{}{ \mathrm{G} }
\NewDocumentCommand\LieAlgebraAny{}{ \mathfrak{g} }

\NewDocumentCommand\LieGroupSO{m}{ \mathrm{SO}(#1) }
\NewDocumentCommand\LieAlgebraSO{m}{ \mathfrak{so}(#1) }

\NewDocumentCommand\LieGroupSE{m}{ \mathrm{SE}(#1) }
\NewDocumentCommand\LieAlgebraSE{m}{ \mathfrak{se}(#1) }

\NewDocumentCommand\LieGroupAdjoint{mm}{ \mathrm{Ad}_{#1}\left(#2\right) }
\NewDocumentCommand\LieAlgebraAdjoint{mm}{ \mathrm{ad}_{#1}\left(#2\right) }
\NewDocumentCommand\LieAlgebraGenerator{m}{ \mathbf{G}_{#1} }

\NewDocumentCommand\Wedge{m}{\left(#1\right)^\wedge}
\NewDocumentCommand\Vee{m}{\left(#1\right)^\vee}

\NewDocumentCommand\Matlog{m}{\mathrm{ln}\left(#1\right)}

\NewDocumentCommand\Matexp{m}{\exp\left(#1\right)}

\NewDocumentCommand\Expectation{m}{ \mathbb{E}\left[#1\right] }
\NewDocumentCommand\NormalDistribution{mm}{ \mathcal{N}\left(#1,#2\right) }

\NewDocumentCommand\Zero{}{ \Matrix{0} }
\NewDocumentCommand\Identity{}{ \Matrix{I} }

\NewDocumentCommand\Mean{m}{\bar{#1}}

\NewDocumentCommand\Inv{m}{{#1}^{-1}}
\NewDocumentCommand\Defined{}{\triangleq}
\NewDocumentCommand\T{}{\mathsf{T}}

\NewDocumentCommand\LieGroupGL{m}{ \mathrm{GL}(#1) }

\NewDocumentCommand\LieGroupGal{m}{ \mathrm{Gal}(#1) }

\NewDocumentCommand\LieGroupSGal{m}{ \mathrm{SGal}(#1) }
\NewDocumentCommand\LieAlgebraSGal{m}{ \mathfrak{sgal}(#1) }
\NewDocumentCommand\LieGroupSETwo{m}{ \mathrm{SE_{2}}(#1) }
\NewDocumentCommand\LieAlgebraSETwo{m}{ \mathfrak{se_{\mathrm{2}}}(#1) }

\NewDocumentCommand\CayleyMap{m}{ \mathrm{cay}(#1) }

\NewDocumentCommand\<{}{\mspace{1mu}}
\NewDocumentCommand\nms{}{\mspace{-7mu}}

\makeatletter
\newcommand{\vast}{\bBigg@{3}}
\newcommand{\Vast}{\bBigg@{4}}
\makeatother

\newcommand{\uw}{\Vector{u}^{\wedge}}
\newcommand{\nuw}{\Vector{\nu}^{\wedge}}
\newcommand{\puw}{\Vector{\rho}^{\wedge}}

\NewDocumentCommand\Euclid{}{ \mathbb{E} }

\STARStitle{All About the Galilean Group $\LieGroupSGal{3}$}
\STARSdate{October 28, 2026}      
\STARSauthor{Jonathan Kelly}      
\STARSidentifier{STARS-2023-001}  
\STARSmajorrev{1}                 
\STARSminorrev{43}                

\begin{document}
\maketitle

\begin{abstract}
We consider the Galilean group of affine transformations that preserve absolute time intervals between events in spacetime, as well as spatial distances between simultaneous events.
The special Galilean group, $\LieGroupSGal{3}$, is a 10-dimensional Lie group; we examine the structure of the group and its Lie algebra and discuss the representation of uncertainty on the group manifold.
We also derive closed-form expressions for the $\LieGroupSGal{3}$ exponential and logarithmic maps, the Cayley and inverse Cayley maps, and the left Jacobian and its inverse.
%
Along the way, we mention other groups, including the special orthogonal group, the special Euclidean group, and the group of extended poses, all of which are proper subgroups of the special Galilean group.
We describe the role of time in Galilean relativity and touch on the relationship between temporal and spatial uncertainty.
\end{abstract}

%

\section{Introduction}

The Galilean group is the symmetry group of Galilean relativity: the family of affine transformations that preserve absolute time intervals between \emph{events}, or points in spacetime, and spatial distances between simultaneous events \cite{1971_Levy-Leblond_Galilei}.%
\footnote{The author thanks Prof.\ Robert Mahony for providing this early reference.}\textsuperscript{\normalfont,}\footnote{The group is also sometimes referred to as the \emph{Galilei} group, for example in \cite{1971_Levy-Leblond_Galilei}.}
This is a 10-dimensional group, often denoted $\LieGroupGal{3}$, that is used to describe relationships between inertial reference frames.\footnote{There does not seem to be a standard notational convention for the Galilean group.}
An inertial frame is a reference frame in which Newton's first law of motion holds.
Any nonrotating frame moving at a constant velocity (i.e., undergoing uniform rectilinear motion) relative to an inertial frame is also inertial.
Galilean transformations include spacetime translations, rotations and reflections of spatial coordinates, and Galilean velocity boosts \cite{1971_Levy-Leblond_Galilei,2011_Holm_Geometric_Part_II}.

In this report, we consider the \emph{special} Galilean group $\LieGroupSGal{3}$ and its Lie algebra (for the usual $(3 + 1)$-dimensional spacetime).
Our aims are twofold:
\begin{enumerate}
\setlength{\itemsep}{2pt}
\item to provide a useful (albeit incomplete) reference about the group, and
%
%
\item to illustrate how the group's structure enables uncertainty in position, orientation, velocity, and time to be expressed in a unified way.
\end{enumerate}
Along the way, we review other related groups, including the special orthogonal group, the special Euclidean group, and the group of extended poses \cite{2014_Barrau_Invariant}, each of which is a proper subgroup of the Galilean group.
We highlight the role of time in Galilean relativity and briefly discuss the relationship between spatial and temporal uncertainty.

\section{Preliminaries}
\label{sec:preliminaries}

To begin, we review some preliminaries.
Our notation roughly follows \cite{2024_Barfoot_State}.
Lowercase Latin and Greek letters (e.g., $a$ and $\alpha$) denote scalar variables, while boldface lower- and uppercase letters (e.g., $\Vector{x}$ and $\Matrix{\Theta}$) denote vectors and matrices, respectively.
The $n \times n$ identity matrix is denoted by $\Identity_{n}$ (a departure from \cite{2024_Barfoot_State}) and the $m \times n$ matrix of zeros by $\Zero_{m \times n}$.
When the size is clear from context, we omit the subscript on the matrix $\Zero$.

This report deals with \emph{matrix Lie groups}, which are closed subgroups of the general linear group $\LieGroupGL{n, \Real} \subset \Real^{n \times n}$ of real, invertible matrices.
The group operation is matrix multiplication.
A Lie group is also a smooth manifold.
Each $k$-dimensional Lie group $\LieGroupAny$ has an associated \emph{Lie algebra} $\LieAlgebraAny$, the $k$-dimensional tangent space at the identity element of the group, equipped with a bilinear, skew-symmetric operator $[\cdot, \cdot] \colon \LieAlgebraAny \times \LieAlgebraAny \to \LieAlgebraAny$, called the \emph{Lie bracket}, which satisfies the Jacobi identity.
The Lie bracket describes the local non-commutativity of the Lie group (see \Cref{sec:sgal3_lie_algebra}).
Since the tangent space is a vector space, it has a basis; the basis elements $\{\LieAlgebraGenerator{1}, \dots, \LieAlgebraGenerator{k}\}$ are called the \emph{generators} of the Lie algebra.
Any element of $\LieAlgebraAny$ can therefore be expressed as a linear combination of the generators, that is, by a vector of $k$ real coefficients.

Some other details about groups and manifolds will be useful.
A group \emph{homomorphism} is a map $f \colon \LieGroupAny \to \mathrm{H}$ between two groups $\LieGroupAny$ and $\mathrm{H}$ that preserves the group operation,
\begin{equation*}
f\left(g_{1} \cdot g_{2}\right) =
f\left(g_{1}\right) \cdot f\left(g_{2}\right),
\quad g_{1}, g_{2} \in \LieGroupAny,
\end{equation*}
where the product on the left side is in $\LieGroupAny$ and on the right side is in $\mathrm{H}$.
A group \emph{isomorphism} is a homomorphism that is also bijective.
A \emph{diffeomorphism} is a smooth, bijective map between smooth manifolds whose inverse is also smooth.
A Lie group isomorphism is a group isomorphism that is also a diffeomorphism; it therefore preserves both the algebraic structure and the smooth manifold structure.
Near the identity, a Lie group is diffeomorphic to its Lie algebra, and so the group can often be replaced (locally) by its Lie algebra.
Working with a vector space, rather than a more complicated, curved manifold, is a big win.
This local diffeomorphism is given by the exponential map, $\exp \colon \LieAlgebraAny \to \LieGroupAny$, which takes a neighbourhood of zero in $\LieAlgebraAny$ to a neighbourhood of the identity in $\LieGroupAny$; restricted to these neighbourhoods, its inverse is the logarithmic map, $\ln \colon \LieGroupAny \to \LieAlgebraAny$.

\section{Galilean Relativity and Inertial Reference Frames}
\label{sec:galilean_relativity}

The principle of Galilean relativity states that the laws of mechanics are the same for all observers in uniform motion relative to one another, although they may assign different spatial and temporal coordinates to their observations.
Before introducing the Lie group structure used later in the report, we summarize the geometric framework on which this principle rests: Galilean spacetime, inertial reference frames, the Galilean transformations that relate one inertial frame to another, and the quantities that these transformations preserve.

\subsection{The Structure of Galilean Spacetime}
\label{subsec:spacetime}

Galilean relativity unifies space and time into a single four-dimensional affine space, known as Galilean spacetime, whose points are called \emph{events} \cite{1981_Artz_Classical}.
Time is modelled as a one-dimensional Euclidean space, $\Euclid$, whose points are instants.
All events that occur at a given instant form a three-dimensional Euclidean space, $\Euclid^{3}$.
Time is absolute, in the sense that the instant at which an event occurs does not depend on the observer, although the assigned time coordinate may depend on the choice of time origin.
Consequently, the time interval between any two events is observer-independent.

This structure makes Galilean spacetime a fibre bundle whose base space is absolute time, $\Euclid$, and whose fibre over each instant is a three-dimensional Euclidean space, $\Euclid^{3}$, containing the events at that instant \cite{2005_Penrose_Road}.
In this view, spacetime is a one-parameter family of Euclidean spaces indexed by time.
%
Two events are \emph{simultaneous} if they lie in the same fibre; because the fibres are defined without reference to any observer, simultaneity is absolute \cite{1989_Arnold_Mathematical,2005_Bhand_Rigid}.
A time interval is defined between any two events, whereas an observer-independent spatial distance is defined only between simultaneous events \cite[Chapter~6]{2014_Godinho_Introduction}.
There is no natural correspondence between the points of different fibres, and hence no observer-independent notion of two events at different instants occurring at the same place \cite{2012_Maudlin_Philosophy}.
%
%

\subsection{Inertial Frames and Inertial Motion}
\label{subsec:inertial_frames}

An inertial reference frame may be thought of as a standard Cartesian spatial frame (i.e., an orthogonal triad of coordinate axes) whose motion is uniform and whose orientation remains fixed.
A particular \emph{coordinate} representation also requires the choice of a time origin and a spatial origin.

There is no privileged inertial frame, and hence uniform translational motion is relative.
However, as noted previously, time is absolute in Galilean relativity---all observers agree on a single, uniformly flowing time, unaffected by motion.
Their time coordinates may differ because they choose different time origins.
Moreover, simultaneity is absolute: if two events are simultaneous in one inertial frame, they remain simultaneous in all inertial frames
\cite{1989_Arnold_Mathematical,2005_Bhand_Rigid}; see \Cref{subsec:geometric_invariants} for more details on simultaneity.

The path of a body through Galilean spacetime is called its \emph{worldline}.
This is a curve whose parameter is absolute time and whose values lie in the successive spatial fibres associated with each moment.
A worldline is straight precisely when the motion is inertial.
Any deviation from a straight worldline indicates that the body is accelerating.

\subsection{Galilean Transformations}
\label{subsec:galilean_transformations}

Inertial reference frames are related by \emph{Galilean transformations}, composed of spatial rotations, spacetime translations, and Galilean boosts.
These operations generate the full set of admissible coordinate changes between inertial frames.%
\footnote{Throughout, we consider only proper spatial rotations---hence the $\mathrm{S}$ in $\LieGroupSGal{3}$---and also exclude time reversal.}
We first consider how each elementary operation acts on an event, before presenting the group structure and composition of transformations in \Cref{sec:sgal3_group}.

For the developments that follow later on, it will be convenient to write the coordinates of an event as a five-element homogeneous column.
These columns form the set
\begin{equation}
\label{eqn:event_definition}
\mathcal{E} \Defined
\begin{Bmatrix}
\Vector{p} =
\bbm
\Vector{x} \\
t \\
1
\ebm
\ \vast\vert \
\Vector{x} =
\bbm
x \\
y \\
z
\ebm \in \Real^{3},
t \in \Real
\end{Bmatrix}.
\end{equation}
Homogeneous coordinates allow Galilean transformations to act on events by matrix multiplication, as shown in \Cref{sec:sgal3_group}.%
\footnote{An event and its coordinates are not the same thing, but we will often treat them as synonymous.}
We write event coordinates $(\Vector{x}, t)$ as elements of $\Real^{3} \times \Real$ to emphasize the distinct roles of space and time (see \Cref{subsec:geometric_invariants}).

\subsubsection{Spatial Rotations}
\label{subsubsec:rotations}

A spatial rotation is a linear transformation of spatial coordinates and is represented by an element of the \emph{special orthogonal group},
\begin{equation}
\label{eqn:SO3_definition}
\LieGroupSO{3} \Defined
\left\{
\Matrix{C} \in \Real^{3 \times 3} \,\Big|\,
\Matrix{C}\<\Matrix{C}^{\T} = \Identity_{3}, \Determinant{\Matrix{C}} = 1
\right\}.
\end{equation}
A rotation acts only on the spatial coordinates of an event,
\begin{equation}
(\Vector{x}, t) \mapsto (\Matrix{C}\<\Vector{x}, t).
\end{equation}
Because $\Matrix{C}$ is orthogonal, Euclidean distances are preserved.\footnote{Since $\Determinant{\Matrix{C}} = +1$, we consider proper rotations only, which preserve spatial handedness.}

Later, we will require the Lie algebra of $\LieGroupSO{3}$, denoted $\LieAlgebraSO{3}$, which consists of the $3 \times 3$ skew-symmetric matrices \cite[Chapter~4]{2005_Selig_Geometric},
\begin{equation}
\label{eqn:so3_definition}
\LieAlgebraSO{3} \Defined
\left\{
\Matrix{\Phi} = \Vector{\phi}^{\wedge} \in \Real^{3 \times 3}
\,\Big|\, \Vector{\phi} \in \Real^{3}
\right\}.
\end{equation}
The linear `wedge' operator $\Wedge{\cdot} \colon \Real^{3} \to \LieAlgebraSO{3}$ maps a vector to its corresponding skew-symmetric matrix,
\begin{equation}
\Vector{\phi}^{\wedge} =
\bbm
\phi_{1} \\
\phi_{2} \\
\phi_{3}
\ebm^{\wedge} =
\bbm
0 & -\phi_{3} & \phi_{2} \\
\phi_{3} & 0 & -\phi_{1} \\
-\phi_{2} & \phi_{1} & 0
\ebm.
\end{equation}
The `vee' operator $\Vee{\cdot}$ recovers $\Vector{\phi}$ from $\Matrix{\Phi}$,
\begin{equation*}
\Matrix{\Phi} = \Vector{\phi}^{\wedge}
\quad \longleftrightarrow \quad
\Vector{\phi} = \Matrix{\Phi}^{\vee}.
\end{equation*}

\subsubsection{Spacetime Translations}
\label{subsubsec:translations}

The coordinates of an event may be translated in space and time by the pair $(\Vector{r}, \tau) \in \Real^{3} \times \Real$,
\begin{equation}
(\Vector{x}, t) \mapsto
(\Vector{x} + \Vector{r}, t + \tau).
\end{equation}
A spatial translation is \emph{isochronous}, in the sense that translating the spatial coordinates of an event does not alter its time coordinate.
A time translation shifts all events uniformly along the absolute time axis.
%

\subsubsection{Galilean Boosts}
\label{subsubsec:boosts}

Inertial reference frames may undergo uniform rectilinear motion with respect to one another.
A \emph{Galilean boost} describes this relative motion.
For a velocity $\Vector{v} \in \Real^{3}$, the action of a boost on an event $(\Vector{x}, t)$ is
\begin{equation}
(\Vector{x}, t) \mapsto (\Vector{x} + \Vector{v}t, t).
\end{equation}
Only relative velocities matter---just as Galilean spacetime has no preferred origin, it also has no preferred state of rest or uniform motion \cite{2012_Maudlin_Philosophy}.

\begin{infobox}[t!]
\begin{mdframed}
\textbf{Why is Galilean Spacetime Affine?}
\vspace{0.5\baselineskip}
\par\noindent Galilean spacetime has an affine structure, rather than a vector space structure.
What, exactly, does this mean?
Quoting from Artz \cite{1981_Artz_Classical}: ``The essential difference between vector spaces and flat spaces is that the former have preferred points, namely their zeros, while the latter do not.
(Thus the latter are more suitable as mathematical models of physical spaces and space-times.)''

\par\hspace{12pt} Stated in another way, Galilean spacetime has no preferred origin, that is, no privileged event that should be treated as `zero' (although this can be imposed, if desired).
Two events can be subtracted to obtain a translation vector between them; this operation does not require the choice of an origin.
However, events themselves cannot be `added' in a meaningful way \cite{2002_Bhand_Rigid}.
%
\end{mdframed}
\end{infobox}

\subsection{Geometric Invariants}
\label{subsec:geometric_invariants}

What quantities are preserved, or remain \emph{invariant}, under special Galilean transformations?
The following quantities are invariant:
\begin{itemize}
\setlength{\itemsep}{2pt}
\item The time interval $t_{2} - t_{1}$ between any two events $(\Vector{x}_{1}, t_{1})$ and $(\Vector{x}_{2}, t_{2})$ is invariant.
\item The distance in space \emph{at the same time} between any two events $(\Vector{x}_{1}, t_{0})$ and $(\Vector{x}_{2}, t_{0})$, $\Norm{\Vector{x}_{2} - \Vector{x}_{1}}_{2}$, is invariant.
\item The handedness of space is invariant, since only proper rotations (with determinant $+1$) are allowed.
\end{itemize}
Because time intervals are invariant, simultaneity is also preserved \cite[Chapter~6]{2014_Godinho_Introduction}: two events $(\Vector{x}_{1}, t_{1})$ and $(\Vector{x}_{2}, t_{2})$ that are simultaneous ($t_{2} - t_{1} = 0$) in one inertial frame are simultaneous in every frame \cite{1989_Arnold_Mathematical,2002_Bhand_Rigid}.

Notably, Galilean spacetime is not equipped with the standard Euclidean metric on $\Real^{4}$.
Time intervals and spatial distances between simultaneous events are measured separately and cannot be `combined' into a single Galilean-invariant Euclidean distance.
This reflects the fibre bundle structure of Galilean spacetime described in \Cref{subsec:spacetime}.%
\footnote{In special relativity, time intervals and spatial distances do combine into a single invariant interval, defined by the Minkowski metric.}

\section{The Matrix Lie Group $\LieGroupSGal{3}$}
\label{sec:sgal3_group}

We consider the connected component at the identity of $\LieGroupGal{3}$, denoted by $\LieGroupSGal{3}$.
Elements of this special Galilean group can be written as $5 \times 5$ matrices,
\begin{equation}
\label{eqn:SGal3_definition}
\LieGroupSGal{3} \Defined
\begin{Bmatrix}
\Matrix{F} =
\bbm
\Matrix{C} & \Vector{v} & \Vector{r} \\
\Zero & 1 & \tau \\
\Zero & 0 & 1
\ebm
\in \Real^{5 \times 5}
\ \vast\vert \
\Matrix{C} \in \LieGroupSO{3},
\Vector{v} \in \Real^{3},
\Vector{r} \in \Real^{3},
\tau \in \Real
\end{Bmatrix}.
\end{equation}
We use $\Matrix{F} \in \LieGroupSGal{3}$ to denote an element of the special Galilean group.
The inverse of $\Matrix{F}$ is
\begin{equation}
\label{eqn:SGal3_inverse}
\Inv{\Matrix{F}} =
\bbm
 \Transpose{\Matrix{C}} &
-\Transpose{\Matrix{C}}\Vector{v} &
-\Transpose{\Matrix{C}}\left(\Vector{r} - \Vector{v}\tau\right) \\
 \Zero & 1 & -\tau \\
 \Zero & 0 & 1
\ebm,
\end{equation}
such that $\Matrix{F}\Inv{\Matrix{F}}\! = \Identity_{5}$.
These matrices form a subgroup of $\LieGroupGL{5, \Real}$, with matrix multiplication as the group operation.
The special Galilean group can be decomposed as $\LieGroupSGal{3} \cong \left(\LieGroupSO{3} \ltimes \Real^{3}\right) \ltimes \left(\Real^{3} \times \Real\right)$, where the inner semidirect product corresponds to rotations and boosts, and the outer semidirect product adds spacetime translations.
We make use of the matrix representation throughout the remainder of the report.
An element $\Matrix{F}$ acts on an event $\Vector{p} \in \mathcal{E}$ by left multiplication,
\begin{equation}
\label{eqn:SGal3_action}
\Matrix{F}\Vector{p} =
\bbm
\Matrix{C} & \Vector{v} & \Vector{r} \\
\Zero & 1 & \tau \\
\Zero & 0 & 1
\ebm
\bbm
\Vector{x} \\
t \\
1
\ebm
=
\bbm
\Matrix{C}\<\Vector{x} + \Vector{v}t + \Vector{r} \\
t + \tau \\
1
\ebm,
\end{equation}
which applies a rotation, a boost, and a spacetime translation, in that order (see \Cref{subsec:galilean_transformations}).

Several subgroups are obtained by restricting the rotation, boost, or translation parameters.
We list a few of them here:
\begin{itemize}
\setlength{\itemsep}{2pt}
\item The \emph{homogeneous} Galilean group is a six-dimensional subgroup ($\Vector{r} = \Zero$, $\tau = 0$).
It is isomorphic to the quotient of $\LieGroupSGal{3}$ by the normal subgroup of spacetime translations \cite{2002_Bhand_Rigid}.
\item The \emph{anisotropic} Galilean group is a seven-dimensional subgroup ($\Matrix{C} = \Identity_{3}$).
\item The \emph{isochronous} Galilean group is a nine-dimensional subgroup ($\tau = 0$).
\end{itemize}

Notably, the isochronous Galilean group has already appeared in the literature, but under a different name.
The group $\LieGroupSETwo{3}$, described initially in \cite{2014_Barrau_Invariant} and called the \emph{group of extended poses} in \cite{2020_Barrau_Mathematical,2022_Brossard_Associating}, is isomorphic to the isochronous Galilean group under the identification of linear velocity with Galilean boosts.

\section{The Lie Algebra $\LieAlgebraSGal{3}$}
\label{sec:sgal3_lie_algebra}

The set of all tangent vectors at the identity element of $\LieGroupSGal{3}$ defines its Lie algebra $\LieAlgebraSGal{3}$.
This tangent space is a 10-dimensional real vector space, matching the dimension of the group itself.
Elements of $\LieAlgebraSGal{3}$ can be written as $5 \times 5$ matrices.

Consider a smooth curve on $\LieGroupSGal{3}$ parameterized by the real variable $s$ (rather than $t$ for `time,' which would be ambiguous in this case).
Let $\Matrix{F}(0) = \Identity_{5}$.
The tangent vector $\mathring{\Matrix{F}}(s)$ lies in the tangent space at $\Matrix{F}(s)$.
Here, we use the symbol $\mathring{\left(\cdot\right)}$ to indicate that the derivative is with respect to the variable $s$ and not the variable $t$.
Left translation by $\Inv{\Matrix{F}}(s)$ maps this tangent vector back to the identity,
%
%
\begin{align}
\Matrix{\Xi} & =
\at{\Inv{\Matrix{F}}(s)\,\mathring{\Matrix{F}}(s)}{s = 0} \notag \\
& =
\at{
\Inv{\bbm
\Matrix{C}(s) &
\Vector{v}(s) &
\Vector{r}(s) \\
\Zero & 1 & \tau(s) \\
\Zero & 0 & 1
\ebm}
\bbm
\mathring{\Matrix{C}}(s) &
\mathring{\Vector{v}}(s) &
\mathring{\Vector{r}}(s) \\
\Zero & 0 & \mathring{\tau}(s) \\
\Zero & 0 & 0
\ebm}{s = 0} \notag \\
& =
\at{\bbm
\Matrix{C}^{\T}(s)\,\mathring{\Matrix{C}}(s) &
\Matrix{C}^{\T}(s)\,\mathring{\Vector{v}}(s) &
\Matrix{C}^{\T}(s)
\bigl(\mathring{\Vector{r}}(s) - \Vector{v}(s)\,\mathring{\tau}(s)\bigr) \\
\Zero & 0 & \mathring{\tau}(s) \\
\Zero & 0 & 0
\ebm^{\vphantom{1}}}{s = 0}. 
\end{align}
At the identity, $s = 0$, $\Matrix{C}(0) = \Transpose{\Matrix{C}}\!(0) = \Identity_{3}$ and $\Vector{v}(0) = \Zero$.
The definition of $\LieAlgebraSGal{3}$ is then
\begin{equation}
\label{eqn:sgal3_definition}
\LieAlgebraSGal{3} \Defined
\begin{Bmatrix}
\Matrix{\Xi} = \hspace{1pt}
\bbm
\Vector{\phi}^{\wedge} & \Vector{\nu} & \Vector{\rho} \\
\Zero & 0 & \iota \\
\Zero & 0 & 0
\ebm
\in \Real^{5 \times 5}
\ \vast\vert \
\Vector{\phi}  \in \Real^{3},
\Vector{\nu}   \in \Real^{3},
\Vector{\rho}  \in \Real^{3},
\iota \in \Real
\end{Bmatrix},
\end{equation}
where $\iota = \mathring{\tau}(0)$, $\Vector{\rho} = \mathring{\Vector{r}}(0)$, $\Vector{\nu} = \mathring{\Vector{v}}(0)$, and $\Vector{\phi}^{\wedge} = \mathring{\Matrix{C}}(0)$ is a skew-symmetric submatrix of the form shown in
\Cref{subsubsec:rotations}.
We overload the $\Wedge{\cdot}$ operator (as done in several texts, e.g., \cite{1994_Murray_Mathematical}) for convenience,
\begin{equation}
\Vector{\xi}^{\wedge} =
\bbm
\Vector{\rho} \\
\Vector{\nu}  \\
\Vector{\phi} \\
\iota
\ebm^{\wedge}=
\bbm
\Vector{\phi}^{\wedge} & \Vector{\nu} & \Vector{\rho} \\
\Zero & 0 & \iota \\
\Zero & 0 & 0
\ebm \in \LieAlgebraSGal{3},
\end{equation}
as a mapping $\Real^{10} \to \LieAlgebraSGal{3}$.\footnote{Possibly confusingly, the Greek letters $\Matrix{\Xi}$ and $\Vector{\xi}$ are used in \cite{2024_Barfoot_State} and elsewhere to represent elements of $\LieAlgebraSE{3}$; we reuse them here for $\LieAlgebraSGal{3}$ because of a lack of suitable alternatives.}
Similarly, we overload the $\Vee{\cdot}$ operator such that
\begin{equation*}
\Vector{\xi}^{\wedge} = \Matrix{\Xi}
\quad \longleftrightarrow \quad
\Matrix{\Xi}^{\vee} = \Vector{\xi}.
\end{equation*}
The reason for the ordering of the variables in the column will become clear later (in \Cref{sec:adjoint}).
Elements of $\LieAlgebraSGal{3}$ can be written as linear combinations of the 10 generators, $\Vector{\xi}^{\wedge} = \sum_{i = 1}^{10} \xi_{i}\LieAlgebraGenerator{i}$, where $\xi_{i}$ is the $i$th entry of $\Vector{\xi}$ and
\smallskip
\begin{gather*}
\LieAlgebraGenerator{1} =
\bbm
0 & 0 & 0 & 0 & 1 \\
0 & 0 & 0 & 0 & 0 \\
0 & 0 & 0 & 0 & 0 \\
0 & 0 & 0 & 0 & 0 \\
0 & 0 & 0 & 0 & 0
\ebm,
\;\;
\LieAlgebraGenerator{2} =
\bbm
0 & 0 & 0 & 0 & 0 \\
0 & 0 & 0 & 0 & 1 \\
0 & 0 & 0 & 0 & 0 \\
0 & 0 & 0 & 0 & 0 \\
0 & 0 & 0 & 0 & 0
\ebm,
\;\;
\LieAlgebraGenerator{3} =
\bbm
0 & 0 & 0 & 0 & 0 \\
0 & 0 & 0 & 0 & 0 \\
0 & 0 & 0 & 0 & 1 \\
0 & 0 & 0 & 0 & 0 \\
0 & 0 & 0 & 0 & 0
\ebm,
\;\;
\LieAlgebraGenerator{4} =
\bbm
0 & 0 & 0 & 1 & 0 \\
0 & 0 & 0 & 0 & 0 \\
0 & 0 & 0 & 0 & 0 \\
0 & 0 & 0 & 0 & 0 \\
0 & 0 & 0 & 0 & 0
\ebm,
\\[4mm]
%
\LieAlgebraGenerator{5} =
\bbm
0 & 0 & 0 & 0 & 0 \\
0 & 0 & 0 & 1 & 0 \\
0 & 0 & 0 & 0 & 0 \\
0 & 0 & 0 & 0 & 0 \\
0 & 0 & 0 & 0 & 0
\ebm,
\;\;\,\quad\quad
\LieAlgebraGenerator{6} =
\bbm
0 & 0 & 0 & 0 & 0 \\
0 & 0 & 0 & 0 & 0 \\
0 & 0 & 0 & 1 & 0 \\
0 & 0 & 0 & 0 & 0 \\
0 & 0 & 0 & 0 & 0
\ebm,
\;\quad\quad
\LieAlgebraGenerator{7} =
\bbm
0 & 0 &  0 & 0 & 0 \\
0 & 0 & -1 & 0 & 0 \\
0 & 1 &  0 & 0 & 0 \\
0 & 0 &  0 & 0 & 0 \\
0 & 0 &  0 & 0 & 0
\ebm,
\\[4mm]
\LieAlgebraGenerator{8} =
\bbm
 0 & 0 & 1 & 0 & 0 \\
 0 & 0 & 0 & 0 & 0 \\
-1 & 0 & 0 & 0 & 0 \\
 0 & 0 & 0 & 0 & 0 \\
 0 & 0 & 0 & 0 & 0
\ebm,
\quad\quad
\LieAlgebraGenerator{9} =
\bbm
0 & -1 & 0 & 0 & 0 \\
1 &  0 & 0 & 0 & 0 \\
0 &  0 & 0 & 0 & 0 \\
0 &  0 & 0 & 0 & 0 \\
0 &  0 & 0 & 0 & 0
\ebm,
\quad\quad
\LieAlgebraGenerator{10} =
\bbm
0 & 0 & 0 & 0 & 0 \\
0 & 0 & 0 & 0 & 0 \\
0 & 0 & 0 & 0 & 0 \\
0 & 0 & 0 & 0 & 1 \\
0 & 0 & 0 & 0 & 0
\ebm.
\;\;\,\, 
\end{gather*}
The matrix representation of the Lie bracket of the elements $\Matrix{\Xi}_{1}, \Matrix{\Xi}_{2} \in \LieAlgebraSGal{3}$ is
\begin{equation}
[\Matrix{\Xi}_{1}, \Matrix{\Xi}_{2}]
=
\Matrix{\Xi}_{1}\,\Matrix{\Xi}_{2} -
\Matrix{\Xi}_{2}\,\Matrix{\Xi}_{1} \in \LieAlgebraSGal{3}.
\end{equation}
See \cite{1999_Marsden_Introduction} and \cite{2011_Holm_Geometric_Part_II} for further details.
More information about the Lie bracket itself is provided in \cite{2005_Selig_Geometric} and an intuitive description is given by Choset et al.\ in \cite[Section~12.1.3]{2005_Choset_Principles}.

\section{The Exponential and Logarithmic Maps}
\label{sec:exp_log_maps}

Having derived the Lie algebra for the Galilean group, the next step is to determine how to move from the vector space $\LieAlgebraSGal{3}$ to the manifold $\LieGroupSGal{3}$ and back.
The exponential map
from $\LieAlgebraSGal{3}$ to $\LieGroupSGal{3}$ and the logarithmic map from $\LieGroupSGal{3}$ to $\LieAlgebraSGal{3}$ allow us to do this \cite{2011_Chirikjian_Stochastic}.
We derive closed-form expressions for these maps next.
More details are provided in \Cref{app:exp_log_maps}.
The exponential map from $\LieAlgebraSGal{3}$ to $\LieGroupSGal{3}$ is
\begin{align}
\exp\left(\Vector{\xi}^{\wedge}\right)
& =
\sum_{n = 0}^{\infty}\frac{1}{n!}
\left(\Vector{\xi}^{\wedge}\right)^{n} =
\sum_{n = 0}^{\infty}\frac{1}{n!}
\left(
\bbm
\Vector{\rho} \\
\Vector{\nu} \\
\Vector{\phi} \\
\iota
\ebm^{\wedge}
\right)^{n} \notag \\[1mm]
& =
\sum_{n = 0}^{\infty}\frac{1}{n!}
\bbm
\Vector{\phi}^{\wedge} & \Vector{\nu} & \Vector{\rho} \\
\Zero & 0 & \iota \\
\Zero & 0 & 0
\ebm^{n} =
\bbm
\Matrix{C} &
\Matrix{D}\<\Vector{\nu} &
\Matrix{D}\Vector{\rho} +
\Matrix{E}\<\Vector{\nu}\iota \\
\Zero & 1 & \iota \\
\Zero & 0 & 1
\ebm,
\label{eqn:SGal3_exp_map_short}
\end{align}
where the matrices $\Matrix{C}$, $\Matrix{D}$, and $\Matrix{E}$ can all be determined in closed form (as shown below).

Consider the axis-angle rotation parameterization $\Vector{\phi} = \phi\<\Vector{u}$, where $\phi = \Norm{\Vector{\phi}}$ is the angle of rotation about the unit axis $\Vector{u} = \Vector{\phi}/\Norm{\Vector{\phi}}$, for $\Vector{\phi} \neq \Zero$.
The matrix $\Matrix{C}$ is obtained via the exponential map from $\LieAlgebraSO{3}$ to $\LieGroupSO{3}$,
\begin{equation}
\label{eqn:SO3_exp_map_short}
\Matrix{C}
=
\exp\left(\phi\<\uw\right)
=
\sum_{n = 0}^{\infty}
\frac{1}{n!}\bigl(\phi\<\uw\bigr)^{n}
=
\Identity_{3} +
\sin\left(\phi\right)\uw +
\bigl(1 - \cos\left(\phi\right)\bigr)\uw\uw,
\end{equation}
which can be derived with the use of an identity given in \Cref{app:identities}.
The result \Cref{eqn:SO3_exp_map_short} is the well-known Rodrigues' rotation formula \cite[Section~2.2]{1994_Murray_Mathematical}.
Notably, the map from $\LieAlgebraSO{3}$ to $\LieGroupSO{3}$ is surjective but not injective: rotation angles that differ by $2\pi$ about the same axis yield the same rotation matrix $\Matrix{C}$.
The remaining matrices $\Matrix{D}$ and $\Matrix{E}$ are
\begin{equation}
\label{eqn:D_exp_map_short}
\Matrix{D}
=
\sum_{n = 0}^{\infty}
\frac{1}{(n + 1)!}\bigl(\phi\<\uw\bigr)^{n}
=
\Identity_{3} +
\left(\frac{1 - \cos\left(\phi\right)}{\phi}\right)\uw +
\left(\frac{\phi - \sin\left(\phi\right)}{\phi}\right)\uw\uw
\end{equation}
and
\begin{equation}
\label{eqn:E_exp_map_short}
\Matrix{E}
=
\sum_{n = 0}^{\infty}
\frac{1}{(n + 2)!}\bigl(\phi\<\uw\bigr)^{n}
=
\frac{1}{2}\Identity_{3} +
\left(\frac{\phi - \sin\left(\phi\right)}{\phi^{2}}\right)\uw +
\left(\frac{\phi^{2} + 2\cos\left(\phi\right) - 2}{2\phi^{2}}\right)\uw\uw.
\end{equation}
For $\phi = 0$, the limiting values are $\Matrix{C} = \Identity_{3}$, $\Matrix{D} = \Identity_{3}$, and $\Matrix{E} = \frac{1}{2}\Identity_{3}$.
Derivations of the matrices $\Matrix{C}$, $\Matrix{D}$, and $\Matrix{E}$ are provided in \Cref{app:exp_log_maps}.

Determining the logarithmic map from $\LieGroupSGal{3}$ to $\LieAlgebraSGal{3}$ requires more work.
From inspection of \Cref{eqn:SO3_exp_map_short,eqn:D_exp_map_short,eqn:E_exp_map_short}, it is clear that we first need to find $\phi$.
To recover the rotation angle, we employ the matrix trace,
\begin{equation}
\label{eqn:angle_from_C}
\phi =
\cos^{-1}\left(\frac{\Trace{\Matrix{C}} - 1}{2}\right).
\end{equation}
The rotation angle is only determined modulo $2\pi$ for a fixed axis; we obtain a unique rotation vector by restricting $\Norm{\Vector{\phi}} < \pi$.
The logarithmic map from $\LieGroupSO{3}$ to $\LieAlgebraSO{3}$ is then
\begin{equation}
\label{eqn:axis_angle_from_C}
\Vector{\phi} =
\Matlog{\Matrix{C}}^{\vee} =
\left(\frac{\phi}{2\sin\left(\phi\right)}
\left(\Matrix{C} - \Matrix{C}^{\T}\right)\right)^{\vee}.
\end{equation}
%

We will also require the inverse of $\Matrix{D}$, which in closed form is
\begin{equation}
\label{eqn:D_inverse_closed_form}
\Inv{\Matrix{D}} =
\Identity_{3} -
\frac{\phi}{2}\uw +
\Biggl(1 - \frac{\phi}{2}\cot\left(\frac{\phi}{2}\right)\Biggr)
\uw\uw.
\end{equation}
For $\phi = 0$, $\Inv{\Matrix{D}} = \Identity_{3}$.
The logarithmic map from $\LieGroupSGal{3}$ to $\LieAlgebraSGal{3}$ can be found by the following procedure: i) set $\iota = \tau$, ii) find $\phi$ from \Cref{eqn:angle_from_C}, $\Vector{\phi}$ from \Cref{eqn:axis_angle_from_C}, set $\Vector{u} = \Vector{\phi}/\phi$, and find $\Inv{\Matrix{D}}$ from \Cref{eqn:D_inverse_closed_form}, iii) compute $\Vector{\nu} = \Inv{\Matrix{D}}\Vector{v}$, and iv) compute $\Vector{\rho} = \Inv{\Matrix{D}} \left(\Vector{r} - \Matrix{E}\<\Vector{\nu}\iota\right)$.
Compactly, the result is
\begin{equation}
\label{eqn:SGal3_log_map_short}
\Vector{\xi}
=
\Matlog{\Matrix{F}}^{\vee}
=
\bbm
\Inv{\Matrix{D}}\left(\Vector{r} - \Matrix{E}\<\Vector{\nu}\iota\right) \\
\Inv{\Matrix{D}}\Vector{v} \\
\Matlog{\Matrix{C}}^{\vee} \\
\tau
\ebm.
\end{equation}
%
%
The exponential map is derived (in a slightly different format and with fewer details) in \cite{2002_Bhand_Rigid}.
Also, these results (for the exponential and logarithmic maps, and also for the adjoint maps, see \Cref{sec:adjoint}) were developed independently in \cite{2019_Fourmy_Absolute}, where the authors describe a group they call the \emph{IMU deltas group} that has the structure of $\LieGroupSGal{3}$.




\section{The Cayley and Inverse Cayley Maps}
\label{sec:cayley_maps}

The exponential map is not the only map from a matrix Lie algebra to the corresponding Lie group.
One well-known alternative is the Cayley map, which replaces the matrix exponential with a rational expression involving matrix inversion \cite{2007_Selig_Cayley}.
The Cayley map can be convenient both algebraically and computationally, particularly when a matrix inverse is easier to evaluate than a matrix exponential.%
\footnote{This is true for $\LieGroupSGal{3}$ even though both maps have closed forms; only the closed form of the exponential map involves trigonometric functions.}

This section presents the Cayley map for $\LieGroupSGal{3}$.%
\footnote{To the best of the author's knowledge, this map has not appeared previously in the literature; it is, however, a straightforward extension of the result for $\LieGroupSE{3}$ \cite{2007_Selig_Cayley}.}
Several scaling conventions for the Cayley map appear in the literature \cite{2001_Iserles_Cayley-Transform,2007_Selig_Cayley}; we follow the widely used convention
\begin{equation}
\label{eqn:SGal3_cayley_map}
\CayleyMap{\Matrix{\Xi}}
\Defined
\left(
\Identity_{5} - \frac{1}{2}\Matrix{\Xi}
\right)^{-1}
\left(
\Identity_{5} + \frac{1}{2}\Matrix{\Xi}
\right),
\quad
\Matrix{\Xi} \in \LieAlgebraSGal{3},
\end{equation}
where the scaling factor is $1/2$.

Using the representation of $\LieAlgebraSGal{3}$ introduced in \Cref{eqn:sgal3_definition}, an element of the Lie algebra has the form
\begin{equation}
\label{eqn:sgal3_cayley_algebra_element}
\Matrix{\Xi}
=
\Vector{\xi}^{\wedge}
=
\bbm
\Vector{\phi}^{\wedge} & \Vector{\nu} & \Vector{\rho} \\
\Zero & 0 & \iota \\
\Zero & 0 & 0
\ebm.
\end{equation}
Because $\Matrix{\Xi}$ is block upper triangular, its Cayley map has the same block structure.
The upper-left block is $\CayleyMap{\Vector{\phi}^{\wedge}}$, given by the usual Cayley map from $\LieAlgebraSO{3}$ to $\LieGroupSO{3}$.
The lower-right $2 \times 2$ block of $\Matrix{\Xi}$ squares to zero, so its Cayley map is
\begin{equation*}
\bbm
1 & \iota \\
0 & 1
\ebm.
\end{equation*}
The time entry of $\CayleyMap{\Matrix{\Xi}}$ is therefore $\iota$, the same as for the exponential map.
The remaining upper-right $3 \times 2$ block is unrestricted in $\LieGroupSGal{3}$, so $\CayleyMap{\Matrix{\Xi}} \in \LieGroupSGal{3}$.

Expanding about $\Matrix{\Xi} = \Zero$ gives
\begin{equation*}
\CayleyMap{\Matrix{\Xi}}
=
\Identity_{5}
+
\Matrix{\Xi}
+
\frac{1}{2}\Matrix{\Xi}^{2}
+
\mathcal{O}\left(\Matrix{\Xi}^{3}\right),
\end{equation*}
so the first-order term of the Cayley map is $\Matrix{\Xi}$ itself, and the first three terms match those of the exponential map.
Another common convention omits the factor of $1/2$; its first-order term is $2\<\Matrix{\Xi}$, so agreement with the exponential map holds only at zeroth order.

Both the Cayley and exponential maps are defined for every $\Matrix{\Xi} \in \LieAlgebraSGal{3}$, but in general
\begin{equation*}
\CayleyMap{\Matrix{\Xi}}
\neq
\Matexp{\Matrix{\Xi}},
\end{equation*}
that is, the same Lie algebra element generally maps to different group elements under the exponential and Cayley maps.

To derive the closed form of the Cayley map for $\LieGroupSGal{3}$, write
\begin{equation}
\label{eqn:SGal3_cayley_group_element}
\CayleyMap{\Matrix{\Xi}}
=
\bbm
\Matrix{C} & \Vector{v} & \Vector{r} \\
\Zero & 1 & \tau \\
\Zero & 0 & 1
\ebm.
\end{equation}
Rather than evaluating the inverse in \Cref{eqn:SGal3_cayley_map} directly, we multiply both sides of \Cref{eqn:SGal3_cayley_map} on the left by $\Identity_{5} - \frac{1}{2}\Matrix{\Xi}$ to obtain
\begin{equation}
\label{eqn:SGal3_cayley_product}
\left(
\Identity_{5} - \frac{1}{2}\Matrix{\Xi}
\right)
\CayleyMap{\Matrix{\Xi}}
=
\Identity_{5} + \frac{1}{2}\Matrix{\Xi}.
\end{equation}
Substituting \Cref{eqn:sgal3_cayley_algebra_element,eqn:SGal3_cayley_group_element} into \Cref{eqn:SGal3_cayley_product} gives
\begin{equation}
\bbm
\Identity_{3} - \frac{1}{2}\Vector{\phi}^{\wedge}
&
-\frac{1}{2}\Vector{\nu}
&
-\frac{1}{2}\Vector{\rho}
\\[1mm]
\Zero
&
1
&
-\frac{1}{2}\iota
\\[1mm]
\Zero
&
0
&
1
\ebm
\bbm
\Matrix{C}
&
\Vector{v}
&
\Vector{r}
\\[1mm]
\Zero
&
1
&
\tau
\\[1mm]
\Zero
&
0
&
1
\ebm
=
\bbm
\Identity_{3} + \frac{1}{2}\Vector{\phi}^{\wedge}
&
\frac{1}{2}\Vector{\nu}
&
\frac{1}{2}\Vector{\rho}
\\[1mm]
\Zero
&
1
&
\frac{1}{2}\iota
\\[1mm]
\Zero
&
0
&
1
\ebm.
\end{equation}
Setting $\Matrix{A} \Defined \Identity_{3} - \frac{1}{2}\Vector{\phi}^{\wedge}$ and equating the corresponding blocks yields
\begin{equation}
\label{eqn:SGal3_cayley_blocks}
\Matrix{A}\Matrix{C}
=
\Identity_{3} + \frac{1}{2}\Vector{\phi}^{\wedge},
\quad
\Matrix{A}\Vector{v}
=
\Vector{\nu},
\quad
\Matrix{A}\Vector{r}
=
\Vector{\rho} + \frac{1}{2}\Vector{\nu}\tau,
\quad
\tau
=
\iota.
\end{equation}
The rotation block is the usual Cayley map for $\LieGroupSO{3}$ \cite{2007_Selig_Cayley}, and the velocity block has the same form as the translational block of the Cayley map for $\LieGroupSE{3}$.
Substituting $\tau = \iota$ into the position relation in \Cref{eqn:SGal3_cayley_blocks}, the Cayley map for $\LieGroupSGal{3}$ is
\begin{equation}
\label{eqn:SGal3_cayley_closed_form}
\CayleyMap{\Matrix{\Xi}}
=
\bbm
\Inv{\Matrix{A}}
\left(
\Identity_{3} + \frac{1}{2}\Vector{\phi}^{\wedge}
\right)
&
\Inv{\Matrix{A}}\Vector{\nu}
&
\Inv{\Matrix{A}}
\left(
\Vector{\rho}
+
\frac{1}{2}\Vector{\nu}\iota
\right)
\\
\Zero
&
1
&
\iota
\\
\Zero
&
0
&
1
\ebm.
\end{equation}
All blocks of \Cref{eqn:SGal3_cayley_closed_form} depend only on $\Inv{\Matrix{A}}$, which has the closed form
\begin{equation}
\label{eqn:SGal3_cayley_A_inverse}
\Inv{\Matrix{A}}
=
\Identity_{3}
+
\frac{2}{4 + \Vector{\phi}^{\T}\Vector{\phi}}
\left(
\Vector{\phi}^{\wedge}
+
\frac{1}{2}\Vector{\phi}^{\wedge}\Vector{\phi}^{\wedge}
\right),
\end{equation}
so that the rotation block is
\begin{equation}
\label{eqn:SGal3_cayley_rotation}
\Matrix{C}
=
\Identity_{3}
+
\frac{4}{4 + \Vector{\phi}^{\T}\Vector{\phi}}
\left(
\Vector{\phi}^{\wedge}
+
\frac{1}{2}\Vector{\phi}^{\wedge}\Vector{\phi}^{\wedge}
\right).
\end{equation}
Multiplying the rotation relation in \Cref{eqn:SGal3_cayley_blocks} on the left by $\Inv{\Matrix{A}}$ and adding $\Identity_{3} = \Inv{\Matrix{A}}\Matrix{A}$ to both sides gives
\begin{equation}
\Matrix{C} + \Identity_{3}
=
\Inv{\Matrix{A}}
\left(
\Identity_{3} + \frac{1}{2}\Vector{\phi}^{\wedge}
+
\Identity_{3} - \frac{1}{2}\Vector{\phi}^{\wedge}
\right)
=
2\Inv{\Matrix{A}},
\end{equation}
which is the form of the identity in \cite{2007_Selig_Cayley} for the scaling used here.
Replacing $\Inv{\Matrix{A}}$ with $\frac{1}{2}(\Matrix{C} + \Identity_{3})$ in \Cref{eqn:SGal3_cayley_closed_form} expresses the Cayley map for $\LieGroupSGal{3}$ in terms of the rotation matrix,
\begin{equation}
\label{eqn:SGal3_cayley_closed_form_rotation}
\CayleyMap{\Matrix{\Xi}}
=
\bbm
\Matrix{C}
&
\frac{1}{2}\left(\Matrix{C} + \Identity_{3}\right)\Vector{\nu}
&
\frac{1}{2}\left(\Matrix{C} + \Identity_{3}\right)
\left(
\Vector{\rho}
+
\frac{1}{2}\Vector{\nu}\iota
\right)
\\
\Zero
&
1
&
\iota
\\
\Zero
&
0
&
1
\ebm.
\end{equation}
In both forms, the position block contains the additional term $\frac{1}{2}\Vector{\nu}\iota$, which results from the coupling between boosts and time translations in $\LieGroupSGal{3}$.

The inverse Cayley map recovers $\Matrix{\Xi}$ from a group element with the blocks in \Cref{eqn:SGal3_cayley_group_element}.
The time component follows directly from the relation in \Cref{eqn:SGal3_cayley_blocks}, $\iota = \tau$.
Rearranging the rotation relation in \Cref{eqn:SGal3_cayley_blocks} gives
\begin{equation*}
\Matrix{C} - \Identity_{3}
=
\frac{1}{2}\Vector{\phi}^{\wedge}
\left(
\Matrix{C} + \Identity_{3}
\right),
\end{equation*}
so that
\begin{equation*}
\Vector{\phi}^{\wedge}
=
2
\left(
\Matrix{C} - \Identity_{3}
\right)
\left(
\Matrix{C} + \Identity_{3}
\right)^{-1},
\end{equation*}
which is the inverse Cayley map for $\LieGroupSO{3}$ \cite{2007_Selig_Cayley}.
This map is defined whenever $\Matrix{C} + \Identity_{3}$ is invertible, that is, for rotations with angle not equal to $\pi$.
Equivalently, without a matrix inverse,
\begin{equation*}
\Vector{\phi}
=
\frac{2}{1 + \Trace{\Matrix{C}}}
\left(
\Matrix{C} - \Matrix{C}^{\T}
\right)^{\vee}.
\end{equation*}
With $\Vector{\phi}$ known, $\Matrix{A}$ follows from its definition, and the velocity and position relations in \Cref{eqn:SGal3_cayley_blocks} give
\begin{equation*}
\Vector{\nu} = \Matrix{A}\Vector{v},
\quad
\Vector{\rho}
=
\Matrix{A}\Vector{r}
-
\frac{1}{2}\Vector{\nu}\tau.
\end{equation*}

\section{The Adjoint Map and the Adjoint Representation}
\label{sec:adjoint}

Consider a group $\LieGroupAny$ and two elements $a, g \in \LieGroupAny$.
The \emph{conjugation} of $a$ by $g$ is $a \mapsto gag^{-1}$.
The element $gag^{-1}$ is called the \emph{conjugate} of $a$ by $g$.
For a Lie group, the differential of the conjugation map at the identity gives a linear map on the Lie algebra, called the \emph{adjoint map},
$\mathrm{Ad}_{g} \colon \LieAlgebraAny \to \LieAlgebraAny$.
In the context of the Galilean group, conjugation can be considered as a transformation between local and global frames (more on this below).

Frequently, it is useful to express the same tangent vector at a group element using either local or global coordinates.
The adjoint map relates these local and global descriptions in the Lie algebra at the identity.
The linear action of a group on a vector space that preserves the group operation is called a \emph{representation} of the group; the collection of adjoint maps $\mathrm{Ad}_{g}$, one for each $g \in \LieGroupAny$, forms the \emph{adjoint representation}.
To derive the adjoint map for $\LieGroupSGal{3}$, we follow \cite[Section~II.F]{2021_Sola_Micro},
\begin{align*}
\exp\left(
\bigl(\LieGroupAdjoint{\Matrix{F}}{\Vector{\xi}}\bigr)^{\wedge}
\right)\Matrix{F}
& =
\Matrix{F}\exp\left(\Vector{\xi}^{\wedge}\right) \\[1mm]
\exp\left(
\bigl(\LieGroupAdjoint{\Matrix{F}}{\Vector{\xi}}\bigr)^{\wedge}
\right)
& =
\Matrix{F}\exp\left(\Vector{\xi}^{\wedge}\right)\Inv{\Matrix{F}} \\[1mm]
\LieGroupAdjoint{\Matrix{F}}{\Vector{\xi}}
& =
\left(\Matrix{F}\Vector{\xi}^{\wedge}\Inv{\Matrix{F}}\right)^{\vee}.
\end{align*}
The final expression is linear in $\Vector{\xi}$, allowing $\mathrm{Ad}_{\Matrix{F}}$ to be written as a $10 \times 10$ matrix,
%
%
%
\begin{align}
\LieGroupAdjoint{\Matrix{F}}{\Vector{\xi}}
& =
\left(
\Matrix{F}\Vector{\xi}^{\wedge}\Inv{\Matrix{F}}
\right)^{\vee} \notag \\
& =
\left(
\bbm
\Matrix{C} & \Vector{v} & \Vector{r} \\
\Zero & 1 & \tau \\
\Zero & 0 & 1
\ebm
\bbm
\Vector{\phi}^{\wedge} & \Vector{\nu} & \Vector{\rho} \\
\Zero & 0 & \iota \\
\Zero & 0 & 0
\ebm
\bbm
\Transpose{\Matrix{C}} &
-\Transpose{\Matrix{C}}\Vector{v} &
-\Transpose{\Matrix{C}}\left(\Vector{r} - \Vector{v}\tau\right) \\
\Zero & 1 & -\tau \\
\Zero & 0 & 1
\ebm
\right)^{\vee} \notag \\[2mm]
& =
\left(
\bbm
\Matrix{C}\Vector{\phi}^{\wedge}\Transpose{\Matrix{C}} &
\Matrix{C}\Vector{\nu} -
\Matrix{C}\Vector{\phi}^{\wedge}\Transpose{\Matrix{C}}\Vector{v} &
\Matrix{C}\Vector{\rho} -
\Matrix{C}\Vector{\nu}\tau +
\Vector{v}\iota -
\Matrix{C}\Vector{\phi}^{\wedge}\Transpose{\Matrix{C}}\Vector{r} +
\Matrix{C}\Vector{\phi}^{\wedge}\Transpose{\Matrix{C}}\Vector{v}\tau \\
\Zero & 0 & \iota \\
\Zero & 0 & 0
\ebm
\right)^{\vee} \notag \\[2mm]
& =
\left(
\bbm
\left(\Matrix{C}\Vector{\phi}\right)^{\wedge} &
\Matrix{C}\Vector{\nu} +
\Vector{v}^{\wedge}\Matrix{C}\Vector{\phi} &
\Matrix{C}\Vector{\rho} -
\Matrix{C}\Vector{\nu}\tau +
\Vector{v}\iota +
\Vector{r}^{\wedge}\Matrix{C}\Vector{\phi} -
\Vector{v}^{\wedge}\Matrix{C}\Vector{\phi}\<\tau \\
\Zero & 0 & \iota \\
\Zero & 0 & 0
\ebm
\right)^{\vee} \notag \\[2mm]
& =
\bbm
\Matrix{C}\Vector{\rho} -
\Matrix{C}\Vector{\nu}\tau +
\Vector{v}\iota +
\Vector{r}^{\wedge}\Matrix{C}\Vector{\phi} -
\Vector{v}^{\wedge}\Matrix{C}\Vector{\phi}\<\tau \\
\Matrix{C}\Vector{\nu} +
\Vector{v}^{\wedge}\Matrix{C}\Vector{\phi} \\
\Matrix{C}\Vector{\phi} \\
\iota
\ebm
=
\bbm
\Matrix{C} & -\Matrix{C}\tau &
\left(\Vector{r} - \Vector{v}\tau\right)^{\wedge}\Matrix{C} & \Vector{v} \\
\Zero & \Matrix{C} & \Vector{v}^{\wedge}\Matrix{C} & \Zero \\
\Zero & \Zero & \Matrix{C} & \Zero \\
\Zero & \Zero & \Zero & 1
\ebm
\bbm
\Vector{\rho} \\
\Vector{\nu} \\
\Vector{\phi} \\
\iota
\ebm.
\label{eqn:SGal3_adjoint_derivation}
\end{align}
Here, we have made use of the identity
\begin{equation}
\Matrix{C}\<\Vector{x}^{\wedge}\Transpose{\Matrix{C}} =
\left(\Matrix{C}\<\Vector{x}\right)^{\wedge}
\end{equation}
for any $\Matrix{C} \in \LieGroupSO{3}$ and $\Vector{x} \in \Real^{3}$.
The adjoint matrix is therefore
\begin{equation}
\label{eqn:SGal3_adjoint_definition}
\mathrm{Ad}_{\Matrix{F}} =
\bbm
\Matrix{C} & -\Matrix{C}\tau &
\left(\Vector{r} - \Vector{v}\tau\right)^{\wedge}\Matrix{C} & \Vector{v} \\
\Zero & \Matrix{C} & \Vector{v}^{\wedge}\Matrix{C} & \Zero \\
\Zero & \Zero & \Matrix{C} & \Zero \\
\Zero & \Zero & \Zero & 1
\ebm
\in \Real^{10 \times 10}.
\end{equation}
The final form of \Cref{eqn:SGal3_adjoint_derivation} reveals the reason for stacking the components of $\Vector{\xi}$ in the order specified in \Cref{sec:sgal3_lie_algebra}: the adjoint has a block upper triangular structure, with three copies of $\Matrix{C}$ along the diagonal.
This structure simplifies operations such as inversion and determinant evaluation.

The Lie algebra $\LieAlgebraSGal{3}$ has its own adjoint representation, defined through the Lie bracket.
For a fixed $\Matrix{\Xi}$, this is the linear map $\mathrm{ad}_{\Matrix{\Xi}} \colon \LieAlgebraSGal{3} \to \LieAlgebraSGal{3}$.\footnote{The lowercase $\mathrm{ad}$ notation is used to distinguish the Lie algebra adjoint from the Lie group adjoint, $\mathrm{Ad}$.}
To determine its form, we begin with the Lie bracket,
\begin{align}
\LieAlgebraAdjoint{\Matrix{\Xi}_{1}}{\Matrix{\Xi}_{2}}
& =
\left(
\Matrix{\Xi}_{1}\,\Matrix{\Xi}_{2} -
\Matrix{\Xi}_{2}\,\Matrix{\Xi}_{1}
\right)^{\vee} \notag \\
& =
\left(
\bbm
\Vector{\phi}_{1}^{\wedge} & \Vector{\nu}_{1} & \Vector{\rho}_{1} \\
\Zero & 0 & \iota_{1} \\
\Zero & 0 & 0
\ebm
\bbm
\Vector{\phi}_{2}^{\wedge} & \Vector{\nu}_{2} & \Vector{\rho}_{2} \\
\Zero & 0 & \iota_{2} \\
\Zero & 0 & 0
\ebm -
\bbm
\Vector{\phi}_{2}^{\wedge} & \Vector{\nu}_{2} & \Vector{\rho}_{2} \\
\Zero & 0 & \iota_{2} \\
\Zero & 0 & 0
\ebm
\bbm
\Vector{\phi}_{1}^{\wedge} & \Vector{\nu}_{1} & \Vector{\rho}_{1} \\
\Zero & 0 & \iota_{1} \\
\Zero & 0 & 0
\ebm
\right)^{\vee} \notag \\[2mm]
& =
\left(
\bbm
\Vector{\phi}_{1}^{\wedge}\Vector{\phi}_{2}^{\wedge} -
\Vector{\phi}_{2}^{\wedge}\Vector{\phi}_{1}^{\wedge} &
\Vector{\phi}_{1}^{\wedge}\Vector{\nu}_{2} -
\Vector{\phi}_{2}^{\wedge}\Vector{\nu}_{1} &
\Vector{\phi}_{1}^{\wedge}\Vector{\rho}_{2} + \Vector{\nu}_{1}\iota_{2} -
\Vector{\phi}_{2}^{\wedge}\Vector{\rho}_{1} - \Vector{\nu}_{2}\iota_{1} \\
\Zero & 0 & 0 \\
\Zero & 0 & 0
\ebm
\right)^{\vee} \notag \\[2mm]
& =
\bbm
\Vector{\phi}_{1}^{\wedge}\Vector{\rho}_{2} + \Vector{\nu}_{1}\iota_{2} -
\Vector{\phi}_{2}^{\wedge}\Vector{\rho}_{1} - \Vector{\nu}_{2}\iota_{1} \\[1mm]
\Vector{\phi}_{1}^{\wedge}\Vector{\nu}_{2} -
\Vector{\phi}_{2}^{\wedge}\Vector{\nu}_{1} \\[1mm]
\left(
\Vector{\phi}_{1}^{\wedge}\Vector{\phi}_{2}^{\wedge} -
\Vector{\phi}_{2}^{\wedge}\Vector{\phi}_{1}^{\wedge}
\right)^{\vee} \\[1mm]
0
\ebm
=
\bbm
\Vector{\phi}_{1}^{\wedge} &
-\Identity_{3}\iota_{1} &
\Vector{\rho}_{1}^{\wedge} &
\Vector{\nu}_{1} \\[1mm]
\Zero & \Vector{\phi}_{1}^{\wedge} & \Vector{\nu}_{1}^{\wedge} & \Zero \\[1mm]
\Zero & \Zero & \Vector{\phi}_{1}^{\wedge} & \Zero \\[1mm]
\Zero & \Zero & \Zero & 0
\ebm
\bbm
\Vector{\rho}_{2} \\[1mm]
\Vector{\nu}_{2} \\[1mm]
\Vector{\phi}_{2} \\[1mm]
\iota_{2}
\ebm.
\label{eqn:sgal3_adjoint_derivation}
\end{align}
Using the same symbol for the map and its matrix representation, the Lie algebra adjoint is
\begin{equation}
\label{eqn:sgal3_adjoint_definition}
\mathrm{ad}_{\Matrix{\Xi}_{1}} =
\bbm
\Vector{\phi}_{1}^{\wedge} &
-\Identity_{3}\iota_{1} &
\Vector{\rho}_{1}^{\wedge} &
\Vector{\nu}_{1} \\[1mm]
\Zero & \Vector{\phi}_{1}^{\wedge} & \Vector{\nu}_{1}^{\wedge} & \Zero \\[1mm]
\Zero & \Zero & \Vector{\phi}_{1}^{\wedge} & \Zero \\[1mm]
\Zero & \Zero & \Zero & 0
\ebm
\in \Real^{10 \times 10}.
\end{equation}
As an alternative, we could have omitted the $\Vee{\cdot}$ operator in \Cref{eqn:SGal3_adjoint_derivation,eqn:sgal3_adjoint_derivation} and kept the results as $5 \times 5$ matrices instead.


%
%

\section{The Differential of the Exponential Map (Jacobian)}
\label{sec:expmap_differential}

Optimization problems on $\LieGroupSGal{3}$ often require the differential of the exponential map, usually called the \emph{Jacobian}.
The \emph{left} Jacobian $\Matrix{J}_{\ell}(\Vector{\xi})$ is the $10 \times 10$ matrix that relates a small change in $\Vector{\xi}$ to the corresponding change in the group element, applied on the left,
\begin{equation}
\label{eqn:left_jacobian_perturbation}
\Matexp{\left(\Vector{\xi} + \delta\Vector{\xi}\right)^{\wedge}}
\approx
\Matexp{\left(\Matrix{J}_{\ell}(\Vector{\xi})\<\delta\Vector{\xi}\right)^{\wedge}}
\Matexp{\Vector{\xi}^{\wedge}},
\end{equation}
to first order in the perturbation $\delta\Vector{\xi}$.
Omitting a very large amount of detail, it can be shown that the left Jacobian acts on the perturbation $\delta\Vector{\xi}$ according to
\begin{equation}
\label{eqn:left_jacobian_definition}
\Matrix{J}_{\ell}(\Vector{\xi})\<\delta\Vector{\xi}
=
\int_{0}^{1}
\LieGroupAdjoint{\<\Matexp{\alpha\<\Vector{\xi}^{\wedge}}}{\delta\Vector{\xi}}
\,d\alpha
=
\sum_{n = 0}^{\infty}
\frac{1}{(n + 1)!}
\mathrm{ad}_{\Vector{\xi}}^{n}\<\delta\Vector{\xi},
\end{equation}
where $\mathrm{ad}_{\Vector{\xi}}$ is the $10 \times 10$ matrix representation of the adjoint action of $\Vector{\xi}^{\wedge}$ on $\LieAlgebraSGal{3}$.
The corresponding \emph{right} Jacobian, which applies the change on the right, is $\Matrix{J}_{r}(\Vector{\xi}) = \Matrix{J}_{\ell}(-\Vector{\xi})$.

%
%

The derivation of the left Jacobian is tedious, but we are able to make use of our results for the exponential map (see \Cref{app:exp_log_maps} and \Cref{app:jacobian}).
The left Jacobian has the following matrix form,
\begin{equation}
\label{eqn:SGal3_left_jacobian_definition}
\Matrix{J}_{\ell}(\Vector{\xi})
=
\bbm
 \Matrix{D} &
-\Matrix{L}\<\iota & 
 \Matrix{N} & 
 \Matrix{E}\<\Vector{\nu} \\[1mm]
\Zero &
\Matrix{D} &
\Matrix{M} &
\Zero \\[1mm]
\Zero & 
\Zero &
\Matrix{D} &
\Zero \\[1mm]
\Zero &
\Zero &
\Zero &
1
\ebm
\in \Real^{10 \times 10}.
\end{equation}
In \Cref{eqn:SGal3_left_jacobian_definition}, the submatrices $\Matrix{D}$, $\Matrix{E}$, and $\Matrix{L}$ depend on $\Vector{\phi}$ only.
The matrices $\Matrix{D}$ and $\Matrix{E}$ are given by \Cref{eqn:D_exp_map_short} and \Cref{eqn:E_exp_map_short}.\footnote{To reduce clutter, we write these matrices without their arguments throughout, for example, $\Matrix{D}$ rather than $\Matrix{D}(\Vector{\phi})$.}
\bigskip

\noindent The matrix $\Matrix{L}$ is
\begin{align}
\Matrix{L}
& =
\sum_{n = 0}^{\infty}
\frac{n + 1}{(n + 2)!}\bigl(\phi\<\uw\bigr)^{n} \notag \\
& =
\frac{1}{2}\Identity_{3} +
\left(
\frac{\sin\left(\phi\right) - \phi\cos\left(\phi\right)}
{\phi^{2}}
\right)
\uw +
\left(
\frac{\phi^{2} +
 2 - 2\phi\sin\left(\phi\right) - 2\cos\left(\phi\right)}
{2\<\phi^{2}}
\right)
\uw\uw.
\label{eqn:L_jacobian_short}
\end{align}
The matrix $\Matrix{M}$ is
\begin{align}
\Matrix{M}
& =
\sum_{n = 0}^{\infty}
\sum_{m = 0}^{\infty}
\frac{1}{(n + m + 2)!}\phi^{n + m}
\left(\uw\right)^{n}
\nuw
\left(\uw\right)^{m} \notag \\[1mm]
& =
\begin{aligned}[t]
\left(\frac{1 - \cos\left(\phi\right)}{\phi^{2}}\right)
\nuw
& +
\left(
\frac{\phi - \sin\left(\phi\right)}{\phi^{2}}
\right)
\left(
\uw
\nuw +
\nuw
\uw
\right) \\[1mm]
& +
\left(
\frac{2 - \phi\sin\left(\phi\right) - 2\cos\left(\phi\right)}{\phi^{2}}
\right)
\uw
\nuw
\uw +
\left(
\frac{2\phi + \phi\cos\left(\phi\right) - 3\sin\left(\phi\right)}{\phi^{2}}
\right)
\uw
\nuw 
\uw
\uw,
\end{aligned}
\end{align}
%
%
which is also part of the left Jacobian of $\LieGroupSE{3}$, but with $\Vector{\rho}$ instead of $\Vector{\nu}$ (see \Cref{app:jacobian} for the derivation).
Lastly, the matrix $\Matrix{N}$ is most easily expressed as the difference of two matrices:
\begin{equation}
\Matrix{N} = \Matrix{N}_{1} - \Matrix{N}_{2}.
\end{equation}
The matrix $\Matrix{N}_{1}$ is
\begin{align}
\Matrix{N}_{1} 
& =
\sum_{n = 0}^{\infty}
\sum_{m = 0}^{\infty}
\frac{1}{(n + m + 2)!}\phi^{n + m}
\left(\uw\right)^{n}
\puw
\left(\uw\right)^{m} \notag \\[1mm]
& =
\begin{aligned}[t]
\left(\frac{1 - \cos\left(\phi\right)}{\phi^{2}}\right)
\puw
& +
\left(
\frac{\phi - \sin\left(\phi\right)}{\phi^{2}}
\right)
\left(
\uw
\puw +
\puw
\uw
\right) \\[1mm]
& +
\left(
\frac{2 - \phi\sin\left(\phi\right) - 2\cos\left(\phi\right)}{\phi^{2}}
\right)
\uw
\puw
\uw +
\left(
\frac{2\phi + \phi\cos\left(\phi\right) - 3\sin\left(\phi\right)}{\phi^{2}}
\right)
\uw
\puw 
\uw
\uw,
\end{aligned}
\label{eqn:N1_jacobian_short}
\end{align}
%
%
which appears (exactly) as part of the Jacobian of $\LieGroupSE{3}$. 
The matrix $\Matrix{N}_{2}$ is
\begin{align}
\Matrix{N}_{2}
& =
\sum_{n = 0}^{\infty}
\sum_{m = 0}^{\infty}
\frac{n + 1}{(n + m + 3)!}\phi^{n + m}
\left(\uw\right)^{n}
\nuw\iota
\left(\uw\right)^{m} \notag \\[1mm]
&
\begin{multlined}[b]
=
\Biggl(
\left(
\frac{2 - \phi\sin\left(\phi\right) -
2\cos\left(\phi\right)}{\phi^{3}}
\right)
\uw +
\left(
\frac{\phi + \phi\cos\left(\phi\right) -
2\sin\left(\phi\right)}{\phi^{3}}
\right)
\uw
\uw
\Biggr)
\nuw\iota \hspace{1.0in} \\[1mm]
+
\left(
\frac{
4\sin\left(\phi\right) -
 \phi^{2}\sin\left(\phi\right) -
4\phi\cos\left(\phi\right)}
{2\<\phi^{3}}
\right)
\uw
\nuw
\uw
\iota \hspace{1.5in} \\[1mm]
\hspace{0.8in}
+
\left(
\frac{
4 +
\phi^{2} +
\phi^{2}\cos\left(\phi\right) -
4\phi\sin\left(\phi\right) -
4\cos\left(\phi\right)
}
{2\<\phi^{3}}
\right)
\uw
\uw
\nuw
\uw
\iota \\[1mm]
+
\nuw\iota
\Biggl(
\left(
\frac{\phi^{2} + 2\cos\left(\phi\right) - 2}{2\<\phi^{3}}
\right)
\uw +
\left(
\frac{\sin\left(\phi\right) - \phi}{\phi^{3}}
\right)
\uw
\uw
\Biggr).
\end{multlined}
\end{align}
To the best of our knowledge, this result for the Jacobian has not appeared before in the literature.\footnote{As of February 25, 2024.}

\section{The Differential of the Logarithmic Map}
\label{sec:logmap_differential}

Optimization problems on $\LieGroupSGal{3}$ may also require the differential of the logarithmic map.
Evaluated at $\Matrix{F} = \Matexp{\Vector{\xi}^{\wedge}}$, this differential is represented by the inverse of the left Jacobian, $\Inv{\Matrix{J}_{\ell}(\Vector{\xi})}$.
For a small left perturbation $\delta\Vector{\xi}$,
\begin{equation}
\label{eqn:inverse_left_jacobian_perturbation}
\Matlog{
\Matexp{\left(\delta\Vector{\xi}\right)^{\wedge}}
\Matexp{\Vector{\xi}^{\wedge}}
}^{\vee}
\approx
\Vector{\xi}
+
\Inv{\Matrix{J}_{\ell}(\Vector{\xi})}
\<\delta\Vector{\xi},
\end{equation}
to first order in $\delta\Vector{\xi}$.
The corresponding inverse right Jacobian is $\Inv{\Matrix{J}_{r}(\Vector{\xi})} = \Inv{\Matrix{J}_{\ell}(-\Vector{\xi})}$.

The full derivation of the inverse left Jacobian is provided in \Cref{app:inverse_jacobian}.
Since $\Matrix{J}_{\ell}$ is block upper triangular, block back-substitution gives an inverse with the same sparsity pattern; collecting the resulting off-diagonal terms into four $3 \times 3$ matrices gives
\begin{equation}
\label{eqn:SGal3_left_jacobian_inverse_definition}
\Inv{\Matrix{J}_{\ell}(\Vector{\xi})}
=
\bbm
\Inv{\Matrix{D}} &
\Matrix{Q} &
-\Matrix{S} &
-\Matrix{P}\Vector{\nu} \\[1mm]
\Zero &
\Inv{\Matrix{D}} &
-\Matrix{R} &
\Zero \\[1mm]
\Zero &
\Zero &
\Inv{\Matrix{D}} &
\Zero \\[1mm]
\Zero &
\Zero &
\Zero &
1
\ebm
\in \Real^{10 \times 10}.
\end{equation}
The four matrices are defined by
\begin{equation*}
\Matrix{P} = \Inv{\Matrix{D}}\Matrix{E},
\quad
\Matrix{Q} = \Inv{\Matrix{D}}\Matrix{L}\Inv{\Matrix{D}}\iota,
\quad
\Matrix{R} = \Inv{\Matrix{D}}\Matrix{M}\Inv{\Matrix{D}},
\quad
\Matrix{S} = \Inv{\Matrix{D}}\left(\Matrix{N} + \Matrix{L}\Inv{\Matrix{D}}\Matrix{M}\iota\right)\Inv{\Matrix{D}}.
\end{equation*}
The diagonal blocks are given by $\Inv{\Matrix{D}}$ from \Cref{eqn:D_inverse_closed_form}, which we repeat here for completeness,
\begin{equation*}
\Inv{\Matrix{D}} =
\Identity_{3} -
\frac{\phi}{2}\uw +
\Biggl(1 - \frac{\phi}{2}\cot\left(\frac{\phi}{2}\right)\Biggr)
\uw\uw.
\end{equation*}
It turns out that the scalar power series corresponding to $\Inv{\Matrix{D}}$ is the generating function for the Bernoulli numbers, $B_{n}$ (see \Cref{app:inverse_jacobian}),
\begin{equation*}
\frac{x}{e^{x} - 1}
=
\sum_{n = 0}^{\infty}
\frac{B_{n}}{n!}x^{n},
\end{equation*}
where we use the convention $B_{1} = -1/2$.
The matrix $\Matrix{P}$ is
\begin{equation*}
\Matrix{P}
=
-
\sum_{n = 1}^{\infty}
\frac{B_{n}}{n!}
\left(\phi\<\uw\right)^{n - 1}
=
\frac{1}{2}\Identity_{3}
-
\Biggl(
\frac{1}{\phi}
-
\frac{1}{2}\cot\left(\frac{\phi}{2}\right)
\Biggr)
\uw.
\end{equation*}
The matrix $\Matrix{Q}$ is
\begin{equation*}
\Matrix{Q}
=
-
\sum_{n = 1}^{\infty}
\frac{B_{n}}{(n - 1)!}
\left(\phi\<\uw\right)^{n - 1}\iota
=
\left(
\frac{1}{2}\Identity_{3}
-
\frac{\phi - \sin\left(\phi\right)}
{2\left(1 - \cos\left(\phi\right)\right)}
\uw
\right)\iota.
\end{equation*}
The matrix $\Matrix{R}$ is
\begin{equation*}
\Matrix{R}
=
\frac{1}{2}\nuw
-
\Biggl(
\frac{1}{\phi}
-
\frac{1}{2}\cot\left(\frac{\phi}{2}\right)
\Biggr)
\left(
\uw\nuw
+
\nuw\uw
\right)
-
\Biggl(
\frac{2}{\phi}
-
\frac{1}{2}\cot\left(\frac{\phi}{2}\right)
-
\frac{\phi}{4}\csc^{2}\left(\frac{\phi}{2}\right)
\Biggr)
\uw\nuw\uw\uw.
\end{equation*}
Lastly, the matrix $\Matrix{S}$ is
\begin{align*}
\Matrix{S}
&
\begin{multlined}[t]
=
\frac{1}{2}\puw
-
\Biggl(
\frac{1}{\phi}
-
\frac{1}{2}\cot\left(\frac{\phi}{2}\right)
\Biggr)
\left(
\uw\puw
+
\puw\uw
\right)
\\[1mm]
\shoveleft[0.25in]{
-
\Biggl(
\frac{2}{\phi}
-
\frac{1}{2}\cot\left(\frac{\phi}{2}\right)
-
\frac{\phi}{4}\csc^{2}\left(\frac{\phi}{2}\right)
\Biggr)
\uw\puw\uw\uw
+
\frac{1}{\phi}
\Biggl(
\frac{1}{\phi}
-
\frac{1}{2}\cot\left(\frac{\phi}{2}\right)
\Biggr)
\nuw\iota
}
\\[1mm]
\hspace{0.5in}
+
\frac{1}{\phi}
\Biggl(
\frac{2}{\phi}
-
\frac{1}{2}\cot\left(\frac{\phi}{2}\right)
-
\frac{\phi}{4}\csc^{2}\left(\frac{\phi}{2}\right)
\Biggr)
\uw\uw\nuw\iota
+
\Biggl(
\frac{\phi}{8}\cot\left(\frac{\phi}{2}\right)\csc^{2}\left(\frac{\phi}{2}\right)
-
\frac{1}{\phi^{2}}
\Biggr)
\uw\nuw\uw\iota.
\end{multlined}
\end{align*}


\section{Uncertainty on $\LieGroupSGal{3}$}
\label{sec:uncertainty}

\begin{figure}[t!]
\centering
\includegraphics[width=0.32\textwidth]{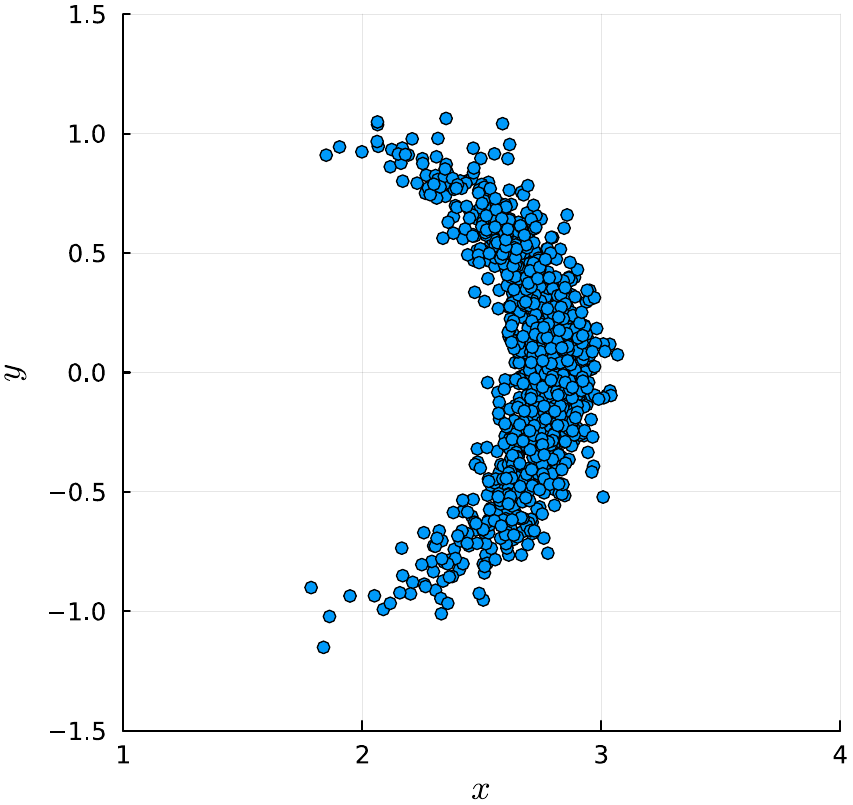}%
\hfill
\includegraphics[width=0.32\textwidth]{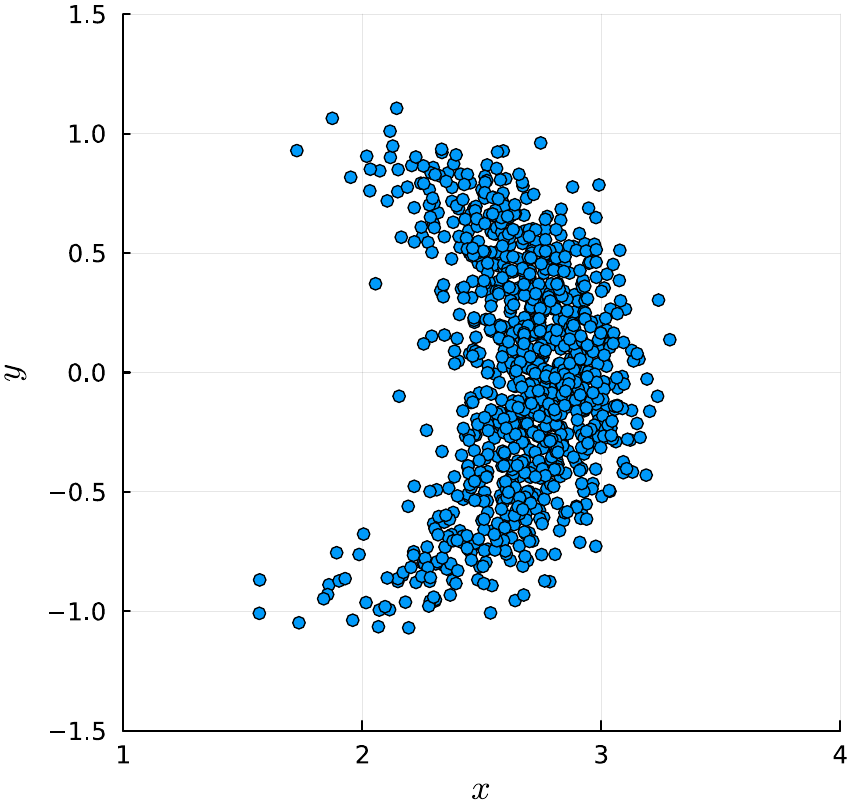}%
\hfill
\includegraphics[width=0.32\textwidth]{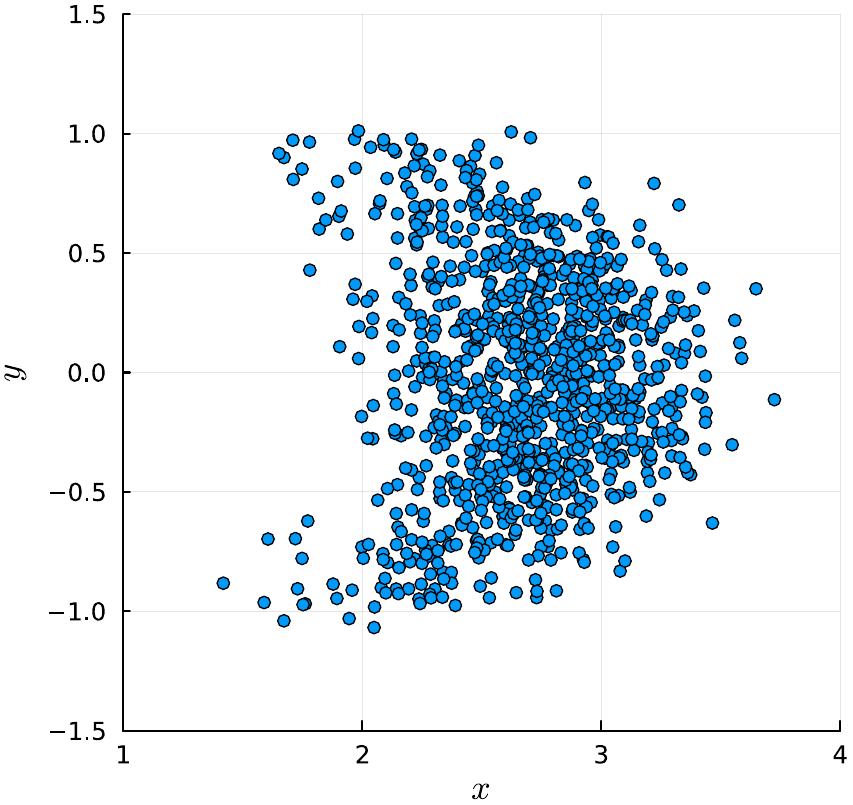}%
\hspace{ 1mm} 
\caption{Visualization of the transformation of an event by a right-perturbed element of $\LieGroupSGal{3}$, projected onto the $x$-$y$ plane. Left: perturbations to the $x$-translation and $z$-rotation components only. Middle: additional (small) perturbation in time. Right: additional (large) perturbation in time. Temporal uncertainty induces a `spread' in the spatial uncertainty. Each plot shows 1,000 samples drawn from a multivariate Gaussian distribution.}
\label{fig:uncertainty}
\vspace{\baselineskip}
\end{figure}

The uncertainty associated with an element of $\LieGroupSGal{3}$ can be expressed as a perturbation in the tangent space.
Following the standard approach, we assume that the perturbation is a vector-valued Gaussian random variable, $\Vector{\xi} \sim \NormalDistribution{\Zero}{\Matrix{\Sigma}}$.
The perturbation can be applied locally (on the right) or globally (on the left),
\begin{equation}
\Matrix{F} = \Mean{\Matrix{F}}\exp\left(\Vector{\xi}^{\wedge}\right)
\quad\text{or}\quad
\Matrix{F} = \exp\left(\Vector{\xi}^{\wedge}\right)\Mean{\Matrix{F}},
\end{equation}
respectively.
If we consider a local (right) perturbation, we can write the covariance of the Gaussian as the expectation
\begin{equation}
\label{eqn:sgal3_covariance}
\Matrix{\Sigma}
\Defined
\Expectation{\Vector{\xi}\Transpose{\Vector{\xi}}}
=
\Expectation{
\Matlog{\Inv{\Mean{\Matrix{F}}}\Matrix{F}}^{\vee} %
\Transpose{\Matlog{\Inv{\Mean{\Matrix{F}}}\Matrix{F}}^{\vee}}}
\in \Real^{10 \times 10}.
\end{equation}
The value of the Galilean group (beyond its use in the physics domain) lies, in part, in the ability to capture spatial and temporal uncertainty in a unified way.
Initial efforts in this direction are described in \cite{2021_Giefer_Uncertainties}, but for $\LieGroupSGal{2}$ only.
Our results are for $\LieGroupSGal{3}$ and include the closed-form maps and Jacobians required for uncertainty propagation on the group.
The examples in \Cref{fig:uncertainty} are limited to 2D projections of 4D events, shown after transformation by an uncertain element of $\LieGroupSGal{3}$.

\section{Closing Remarks}
\label{sec:remarks}

Many problems in physics and engineering involve two or more inertial (or approximately inertial) reference frames that move relative to each other and that may also be offset in time.
The Lie group $\LieGroupSGal{3}$ provides a natural framework for reasoning about such problems and for treating the associated uncertainties.
This short report collects some of the necessary mathematical tools.

\titleformat{\section}{\normalfont\Large\bfseries}{Appendix~\thesection}{0.75em}{}

\appendix

\titleformat{\section}{\normalfont\Large\bfseries}{Appendix~\thesection}{0.75em}{}

\section{The Exponential Map}
\label{app:exp_log_maps}

This appendix provides a derivation of the exponential map for $\LieGroupSGal{3}$ in closed form.
Recall that, for the square matrix $\Matrix{A}$, the matrix exponential is defined by the power series
\begin{equation}
\label{eqn:matrix_exp}
\Matexp{\Matrix{A}}
=
\Identity +
\Matrix{A} +
\frac{1}{2!}\Matrix{A}^{2} +
\frac{1}{3!}\Matrix{A}^{3} +
\cdots
=
\sum_{n = 0}^{\infty}
\frac{1}{n!}\Matrix{A}^{n}.
\end{equation}
The matrix logarithm is defined by the power series
\begin{equation}
\label{eqn:matrix_log}
\ln\left(\Matrix{A}\right)
=
\left(\Matrix{A} - \Identity\right) -
\frac{\left(\Matrix{A} - \Identity\right)^{2}}{2} +
\frac{\left(\Matrix{A} - \Identity\right)^{3}}{3} -
\frac{\left(\Matrix{A} - \Identity\right)^{4}}{4} +
\cdots
=
\sum_{n = 1}^{\infty}
(-1)^{n - 1}\frac{\left(\Matrix{A} - \Identity\right)^{n}}{n},
\end{equation}
which converges when $\Matrix{A}$ is sufficiently close to $\Identity$.
\bigskip

Following \Cref{eqn:matrix_exp}, the exponential map from $\LieAlgebraSGal{3}$ to $\LieGroupSGal{3}$ is
\begin{align}
\exp\left(\Vector{\xi}^{\wedge}\right)
& =
\sum_{n = 0}^{\infty}
\frac{1}{n!}
\bbm
\Vector{\phi}^{\wedge} & \Vector{\nu} & \Vector{\rho} \\
\Zero & 0 & \iota \\
\Zero & 0 & 0
\ebm^{n} \notag \\[0.5mm]
& =
\begin{aligned}[t]
\bbm
\,\Identity_{3} & \Zero & \Zero \\
\Zero & 1 & 0 \\
\Zero & 0 & 1
\ebm
& +
\bbm
\Vector{\phi}^{\wedge} & \Vector{\nu} & \Vector{\rho} \\
\Zero & 0 & \iota \\
\Zero & 0 & 0
\ebm \\[0.5mm]
& +
\frac{1}{2!}
\bbm
\left(\Vector{\phi}^{\wedge}\right)^{2}\!\! &
\Vector{\phi}^{\wedge}\!\Vector{\nu}\!\! &
\Vector{\phi}^{\wedge}\!\Vector{\rho} + \Vector{\nu}\iota \\
\Zero & 0 & 0 \\
\Zero & 0 & 0
\ebm +
\frac{1}{3!}
\bbm
\left(\Vector{\phi}^{\wedge}\right)^{3}\!\! &
\left(\Vector{\phi}^{\wedge}\right)^{2}\!\Vector{\nu}\!\! &
\left(\Vector{\phi}^{\wedge}\right)^{2}\!\Vector{\rho} +
\Vector{\phi}^{\wedge}\Vector{\nu}\iota \\
\Zero & 0 & 0 \\
\Zero & 0 & 0
\ebm +
\cdots
\end{aligned} \notag \\[0.5mm]
& =
\bbm
\Matrix{C} &
\Matrix{D}\<\Vector{\nu} &
\Matrix{D}\Vector{\rho} +
\Matrix{E}\<\Vector{\nu}\iota \\
\Zero & 1 & \iota \\
\Zero & 0 & 1
\ebm.
\label{eqn:SGal3_exp_map_derivation}
\end{align}

To determine the forms of the matrices $\Matrix{C}$, $\Matrix{D}$, and $\Matrix{E}$, we make use of the axis-angle rotation parameterization from \Cref{sec:exp_log_maps} and the identity
\begin{equation*}
\uw\uw\uw = -\uw,
\end{equation*}
when $\Norm{\Vector{u}} = 1$; see \Cref{app:identities} for the derivation.
Any power of $\uw$ greater than two can therefore be expressed in terms of $\uw$ or $\uw\uw$ simply by flipping the minus sign.
Returning to the problem at hand, the upper-left block in \Cref{eqn:SGal3_exp_map_derivation} corresponds to the exponential map from $\LieAlgebraSO{3}$ to $\LieGroupSO{3}$:
\begin{align}
\Matrix{C} =
\exp\left(\Vector{\phi}^{\wedge}\right)
& =
\sum_{n = 0}^{\infty}
\frac{1}{n!}\bigl(\phi\<\uw\bigr)^{n} \notag \\
& =
\Identity_{3} +
\phi\<\uw +
\frac{1}{2!}\phi^{2}\uw\uw +
\frac{1}{3!}\phi^{3}
\underbrace{
\uw\uw\uw}_{-\uw} +
\frac{1}{4!}\phi^{4}
\underbrace{
\uw\uw\uw\uw}_{-\uw \uw} +
\cdots \notag \\
& =
\Identity_{3} +
\left(\phi - \frac{1}{3!}\phi^{3} + \frac{1}{5!}\phi^{5} - \cdots\right)
\uw +
\left(\frac{1}{2!}\phi^{2} - \frac{1}{4!}\phi^{4} + \frac{1}{6!}\phi^{6} - \cdots\right)
\uw\uw \notag \\[2mm]
& =
\Identity_{3} +
\sin\left(\phi\right)\uw +
\bigl(1 - \cos\left(\phi\right)\bigr)
\uw\uw.
\label{eqn:SO3_exp_map}
\end{align}

\noindent The remaining matrices $\Matrix{D}$ and $\Matrix{E}$ are
\begin{align}
\Matrix{D}
& =
\sum_{n = 0}^{\infty}
\frac{1}{(n + 1)!}\bigl(\phi\<\uw\bigr)^{n} \notag \\
& =
\Identity_{3} +
\frac{1}{2!}\phi\<\uw +
\frac{1}{3!}\phi^{2}
\uw\uw +
\frac{1}{4!}\phi^{3}
\uw\uw\uw +
\frac{1}{5!}\phi^{4}
\uw\uw\uw\uw +
\cdots \notag \\
& =
\Identity_{3} +
\left(\frac{1}{2!}\phi - \frac{1}{4!}\phi^{3} + \frac{1}{6!}\phi^{5} - \cdots\right)
\uw +
\left(\frac{1}{3!}\phi^{2} - \frac{1}{5!}\phi^{4} + \frac{1}{7!}\phi^{6} - \cdots\right)
\uw\uw \notag \\
& =
\Identity_{3} +
\left(\frac{1 - \cos\left(\phi\right)}{\phi}\right)
\uw +
\left(\frac{\phi - \sin\left(\phi\right)}{\phi}\right)
\uw\uw
\end{align}
and
\begin{align}
\Matrix{E}
& =
\sum_{n = 0}^{\infty}
\frac{1}{(n + 2)!}\bigl(\phi\<\uw\bigr)^{n} \notag \\
& =
\frac{1}{2}\Identity_{3} +
\frac{1}{3!}\phi\<\uw +
\frac{1}{4!}\phi^{2}
\uw\uw +
\frac{1}{5!}\phi^{3}
\uw\uw\uw +
\frac{1}{6!}\phi^{4}
\uw\uw\uw\uw +
\cdots \notag \\
& =
\frac{1}{2}\Identity_{3} +
\left(\frac{1}{3!}\phi - \frac{1}{5!}\phi^{3} + \frac{1}{7!}\phi^{5} - \cdots\right)
\uw +
\left(\frac{1}{4!}\phi^{2} - \frac{1}{6!}\phi^{4} + \frac{1}{8!}\phi^{6} - \cdots\right)
\uw\uw \notag \\
& =
\frac{1}{2}\Identity_{3} +
\left(\frac{\phi - \sin\left(\phi\right)}{\phi^{2}}\right)
\uw +
\left(\frac{\phi^{2} + 2\cos\left(\phi\right) - 2}{2\phi^{2}}\right)
\uw\uw.
\end{align}
Notably, the closed-form derivation of the exponential map from $\LieAlgebraSGal{3}$ to $\LieGroupSGal{3}$ also yields closed-form expressions for the exponential maps from $\LieAlgebraSE{3}$ to $\LieGroupSE{3}$ and from $\LieAlgebraSETwo{3}$ to $\LieGroupSETwo{3}$ (i.e., the group of extended poses \cite{2022_Brossard_Associating}).\footnote{For cases where the matrix exponential cannot be expressed in closed form, see \cite{2003_Moler_Nineteen} for a useful overview.}
We omit the details, but the results are easily verified.\footnote{This makes sense, of course, since $\LieGroupSE{3}$ and $\LieGroupSETwo{3}$ are both subgroups of $\LieGroupSGal{3}$.}




\section{The Differential of the Exponential Map}
\label{app:jacobian}

The left Jacobian introduced in \Cref{sec:expmap_differential} represents the differential of the exponential map.
Some additional effort is required to determine it in closed form for $\LieGroupSGal{3}$.
In this appendix, we derive the required submatrices.
The matrix $\Matrix{L}$ is
\begin{align}
\Matrix{L}
& =
\sum_{n = 0}^{\infty}
\frac{n + 1}{(n + 2)!}\bigl(\phi\<\uw\bigr)^{n} \notag \\[1mm]
& =
\frac{1}{2}\Identity_{3} +
\frac{2}{3!}\phi\<\uw +
\frac{3}{4!}\phi^{2}
\uw\uw +
\frac{4}{5!}\phi^{3}
\uw\uw\uw +
\frac{5}{6!}\phi^{4}
\uw\uw\uw\uw +
\cdots \notag \\[1mm]
& =
\frac{1}{2}\Identity_{3} +
\left(
\frac{2}{3!}\phi - \frac{4}{5!}\phi^{3} + \frac{6}{7!}\phi^{5} - \cdots
\right)
\uw +
\left(
\frac{3}{4!}\phi^{2} - \frac{5}{6!}\phi^{4} + \frac{7}{8!}\phi^{6} - \cdots
\right)
\uw\uw \notag \\[1mm]
& =
\frac{1}{2}\Identity_{3} +
\left(
\frac{\sin\left(\phi\right) - \phi\cos\left(\phi\right)}
{\phi^{2}}
\right)
\uw +
\left(
\frac{\phi^{2} +
 2 - 2\phi\sin\left(\phi\right) - 2\cos\left(\phi\right)}
{2\<\phi^{2}}
\right)
\uw\uw.
\end{align}
The matrix $\Matrix{M}$ is more complicated.
We begin by writing down the first four terms of the power series,
\begin{align}
\Matrix{M}
& =
\begin{multlined}[t]
\frac{1}{2!}
\nuw +
\frac{1}{3!}
\left(
\Vector{\phi}^{\wedge}\nuw +
\nuw\Vector{\phi}^{\wedge}
\right) +
\frac{1}{4!}
\left(
\Vector{\phi}^{\wedge}
\Vector{\phi}^{\wedge}
\nuw +
\Vector{\phi}^{\wedge}
\nuw
\Vector{\phi}^{\wedge} +
\nuw
\Vector{\phi}^{\wedge}
\Vector{\phi}^{\wedge}
\right) \\
+
\frac{1}{5!}
\left(
\Vector{\phi}^{\wedge}
\Vector{\phi}^{\wedge}
\Vector{\phi}^{\wedge}
\nuw +
\Vector{\phi}^{\wedge}
\Vector{\phi}^{\wedge}
\nuw
\Vector{\phi}^{\wedge} +
\Vector{\phi}^{\wedge}
\nuw
\Vector{\phi}^{\wedge}
\Vector{\phi}^{\wedge} +
\nuw
\Vector{\phi}^{\wedge}
\Vector{\phi}^{\wedge}
\Vector{\phi}^{\wedge}
\right) + \cdots
\end{multlined} \notag \\[2mm]
& =
\begin{multlined}[t]
\frac{1}{2!}
\nuw +
\frac{1}{3!}\phi
\left(
\uw\nuw +
\nuw\uw
\right) +
\frac{1}{4!}\phi^{2}
\left(
\uw
\uw
\nuw +
\uw
\nuw
\uw +
\nuw
\uw
\uw
\right) \\
\hspace*{4.1em} +
\frac{1}{5!}\phi^{3}
\left(
-\uw
 \nuw +
 \uw
 \uw
 \nuw
 \uw +
 \uw
 \nuw
 \uw
 \uw -
 \nuw
 \uw
\right) + \cdots
\end{multlined}
\end{align}
In the second expression above, we have applied \Cref{ident:skew_negative} in \Cref{app:identities} to simplify products involving $\left(\uw\right)^{3} = -\uw$.
Recognizing the emerging pattern, we (eventually) arrive at the closed-form expression
\begin{align*}
\Matrix{M}
& =
\sum_{n = 0}^{\infty}
\sum_{m = 0}^{\infty}
\frac{1}{(n + m + 2)!}\phi^{n + m}
\left(\uw\right)^{n}
\nuw
\left(\uw\right)^{m} \\
&
\begin{aligned}[b]
=
\frac{1}{2}\nuw
& +
\left(
\frac{\phi - \sin\left(\phi\right)}{\phi^{2}}
\right)
\left(
\uw
\nuw +
\nuw
\uw
\right) +
\left(
\frac{\phi - \sin\left(\phi\right)}{\phi}
\right)
\uw
\nuw
\uw \\[1mm]
& +
\left(
\frac{\phi^{2} + 2\cos\left(\phi\right) - 2}{2\<\phi^{2}}
\right)
\left(
 \uw
 \uw
 \nuw +
 \nuw
 \uw
 \uw -
3\<\uw
 \nuw
 \uw
\right) \\[1mm]
& +
\left(
\frac{2\phi + \phi\cos\left(\phi\right) - 3\sin\left(\phi\right)}{2\<\phi^{2}}
\right)
\left(
\uw
\uw
\nuw
\uw +
\uw
\nuw
\uw
\uw
\right),
\end{aligned}
\end{align*}
which is a result originally presented in \cite{2014_Barfoot_Associating} and where we note that $\left(\uw\uw\nuw\uw\uw\right) = - \left(\uw\nuw\uw\right)$, and so on.
Interestingly, we are able to apply two identities from \Cref{app:identities} to further simplify this expression.
Making use of \Cref{ident:skew_products_triple} and \Cref{ident:skew_products_quad} and collecting like terms, we obtain
\begin{align*}
\Matrix{M}
=
\left(\frac{1 - \cos\left(\phi\right)}{\phi^{2}}\right)
\nuw
& +
\left(
\frac{\phi - \sin\left(\phi\right)}{\phi^{2}}
\right)
\left(
\uw
\nuw +
\nuw
\uw
\right) \\[1mm]
& +
\left(
\frac{2 - \phi\sin\left(\phi\right) - 2\cos\left(\phi\right)}{\phi^{2}}
\right)
\uw
\nuw
\uw +
\left(
\frac{2\phi + \phi\cos\left(\phi\right) - 3\sin\left(\phi\right)}{\phi^{2}}
\right)
\uw
\nuw
\uw
\uw,
\end{align*}
a result that may be of independent interest, since this is a more compact expression for part of the Jacobian of $\LieGroupSE{3}$ and $\LieGroupSETwo{3}$.

Finally, we follow the same procedure to find the matrix $\Matrix{N}$ in closed form, by expanding the first five terms in the power series,
\begin{align*}
\Matrix{N}
& =
\begin{multlined}[t]
\frac{1}{2!}
\puw +
\frac{1}{3!}
\left(
\Vector{\phi}^{\wedge}
\puw +
\puw
\Vector{\phi}^{\wedge} -
\nuw\iota
\right) +
\frac{1}{4!}
\left(
 \Vector{\phi}^{\wedge}
 \Vector{\phi}^{\wedge}
 \puw +
 \Vector{\phi}^{\wedge}
 \puw
 \Vector{\phi}^{\wedge} +
 \puw
 \Vector{\phi}^{\wedge}
 \Vector{\phi}^{\wedge} -
2\Vector{\phi}^{\wedge}
 \nuw\iota -
 \nuw
 \Vector{\phi}^{\wedge}\iota
\right) \\
\hspace*{-1em}
+
\frac{1}{5!}
\left(
 \Vector{\phi}^{\wedge}
 \Vector{\phi}^{\wedge}
 \Vector{\phi}^{\wedge}
 \puw +
 \Vector{\phi}^{\wedge}
 \Vector{\phi}^{\wedge}
 \puw
 \Vector{\phi}^{\wedge} +
 \Vector{\phi}^{\wedge}
 \puw
 \Vector{\phi}^{\wedge}
 \Vector{\phi}^{\wedge} +
 \puw
 \Vector{\phi}^{\wedge}
 \Vector{\phi}^{\wedge}
 \Vector{\phi}^{\wedge}
-
3\<\Vector{\phi}^{\wedge}
 \Vector{\phi}^{\wedge}
 \nuw\iota -
2\<\Vector{\phi}^{\wedge}
 \nuw
 \Vector{\phi}^{\wedge}\iota -
 \nuw
 \Vector{\phi}^{\wedge}
 \Vector{\phi}^{\wedge}\iota
\right) \\
+
\frac{1}{6!}
\left(
\Vector{\phi}^{\wedge}
\Vector{\phi}^{\wedge}
\Vector{\phi}^{\wedge}
\Vector{\phi}^{\wedge}
\puw +
\Vector{\phi}^{\wedge}
\Vector{\phi}^{\wedge}
\Vector{\phi}^{\wedge}
\puw
\Vector{\phi}^{\wedge} +
\Vector{\phi}^{\wedge}
\Vector{\phi}^{\wedge}
\puw
\Vector{\phi}^{\wedge}
\Vector{\phi}^{\wedge} +
\Vector{\phi}^{\wedge}
\puw
\Vector{\phi}^{\wedge}
\Vector{\phi}^{\wedge}
\Vector{\phi}^{\wedge} +
\puw
\Vector{\phi}^{\wedge}
\Vector{\phi}^{\wedge}
\Vector{\phi}^{\wedge}
\Vector{\phi}^{\wedge}
\right. \\
\left.
-~
4\<\Vector{\phi}^{\wedge}
 \Vector{\phi}^{\wedge}
 \Vector{\phi}^{\wedge}
 \nuw\iota -
3\<\Vector{\phi}^{\wedge}
 \Vector{\phi}^{\wedge}
 \nuw
 \Vector{\phi}^{\wedge}\iota -
2\<\Vector{\phi}^{\wedge}
 \nuw
 \Vector{\phi}^{\wedge}
 \Vector{\phi}^{\wedge}\iota -
 \nuw
 \Vector{\phi}^{\wedge}
 \Vector{\phi}^{\wedge}
 \Vector{\phi}^{\wedge}\iota
\right) + \cdots
\end{multlined}
\end{align*}
We have seen part of the series before when deriving the matrix $\Matrix{M}$, but with $\nuw$ instead of $\puw$.
We separate $\Matrix{N}$ into two parts,
\begin{align}
\Matrix{N}
& =
\underbrace{
\sum_{n = 0}^{\infty}
\sum_{m = 0}^{\infty}
\frac{1}{(n + m + 2)!}\phi^{n + m}
\left(\uw\right)^{n}
\puw
\left(\uw\right)^{m}
}_{\Matrix{N}_{1}}
-
\underbrace{
\sum_{n = 0}^{\infty}
\sum_{m = 0}^{\infty}
\frac{n + 1}{(n + m + 3)!}\phi^{n + m}
\left(\uw\right)^{n}
\nuw\iota
\left(\uw\right)^{m}
}_{\Matrix{N}_{2}}.
\end{align}
The matrix $\Matrix{N}_{1}$ is
\begin{align*}
\Matrix{N}_{1}
& =
\sum_{n = 0}^{\infty}
\sum_{m = 0}^{\infty}
\frac{1}{(n + m + 2)!}\phi^{n + m}
\left(\uw\right)^{n}
\puw
\left(\uw\right)^{m} \\[1mm]
& =
\begin{aligned}[t]
\left(\frac{1 - \cos\left(\phi\right)}{\phi^{2}}\right)
\puw
& +
\left(
\frac{\phi - \sin\left(\phi\right)}{\phi^{2}}
\right)
\left(
\uw
\puw +
\puw
\uw
\right) \\[1mm]
& +
\left(
\frac{2 - \phi\sin\left(\phi\right) - 2\cos\left(\phi\right)}{\phi^{2}}
\right)
\uw
\puw
\uw +
\left(
\frac{2\phi + \phi\cos\left(\phi\right) - 3\sin\left(\phi\right)}{\phi^{2}}
\right)
\uw
\puw
\uw
\uw.
\end{aligned}
\end{align*}
Finding the closed form of $\Matrix{N}_{2}$ requires several additional steps.
First, we write down the first six terms in the power series to make the pattern (fully) clear,
\begin{align*}
\Matrix{N}_{2}
& =
\begin{multlined}[t]
\frac{1}{3!}\<
\nuw\iota +
\frac{1}{4!}
\left(
2\Vector{\phi}^{\wedge}
 \nuw\iota +
 \nuw
 \Vector{\phi}^{\wedge}\iota
\right) +
\frac{1}{5!}
\left(
3\<\Vector{\phi}^{\wedge}
 \Vector{\phi}^{\wedge}
 \nuw\iota +
2\<\Vector{\phi}^{\wedge}
 \nuw
 \Vector{\phi}^{\wedge}\iota +
 \nuw
 \Vector{\phi}^{\wedge}
 \Vector{\phi}^{\wedge}\iota
\right) \\
+
\frac{1}{6!}
\left(
4\<\Vector{\phi}^{\wedge}
 \Vector{\phi}^{\wedge}
 \Vector{\phi}^{\wedge}
 \nuw\iota +
3\<\Vector{\phi}^{\wedge}
 \Vector{\phi}^{\wedge}
 \nuw
 \Vector{\phi}^{\wedge}\iota +
2\<\Vector{\phi}^{\wedge}
 \nuw
 \Vector{\phi}^{\wedge}
 \Vector{\phi}^{\wedge}\iota +
 \nuw
 \Vector{\phi}^{\wedge}
 \Vector{\phi}^{\wedge}
 \Vector{\phi}^{\wedge}\iota
\right) \\
+
\frac{1}{7!}
\left(
5\<\Vector{\phi}^{\wedge}
 \Vector{\phi}^{\wedge}
 \Vector{\phi}^{\wedge}
 \Vector{\phi}^{\wedge}
 \nuw\iota +
4\<\Vector{\phi}^{\wedge}
 \Vector{\phi}^{\wedge}
 \Vector{\phi}^{\wedge}
 \nuw
 \Vector{\phi}^{\wedge}\iota +
3\<\Vector{\phi}^{\wedge}
 \Vector{\phi}^{\wedge}
 \nuw
 \Vector{\phi}^{\wedge}
 \Vector{\phi}^{\wedge}\iota +
2\<\Vector{\phi}^{\wedge}
 \nuw
 \Vector{\phi}^{\wedge}
 \Vector{\phi}^{\wedge}
 \Vector{\phi}^{\wedge}\iota +
 \nuw
 \Vector{\phi}^{\wedge}
 \Vector{\phi}^{\wedge}
 \Vector{\phi}^{\wedge}
 \Vector{\phi}^{\wedge}\iota
\right) \\
+
\frac{1}{8!}
\left(
6\<\Vector{\phi}^{\wedge}
 \Vector{\phi}^{\wedge}
 \Vector{\phi}^{\wedge}
 \Vector{\phi}^{\wedge}
 \Vector{\phi}^{\wedge}
 \nuw\iota +
5\<\Vector{\phi}^{\wedge}
 \Vector{\phi}^{\wedge}
 \Vector{\phi}^{\wedge}
 \Vector{\phi}^{\wedge}
 \nuw
 \Vector{\phi}^{\wedge}\iota +
4\<\Vector{\phi}^{\wedge}
 \Vector{\phi}^{\wedge}
 \Vector{\phi}^{\wedge}
 \nuw
 \Vector{\phi}^{\wedge}
 \Vector{\phi}^{\wedge}\iota
\right. \\
\left.
+~
3\<\Vector{\phi}^{\wedge}
 \Vector{\phi}^{\wedge}
 \nuw
 \Vector{\phi}^{\wedge}
 \Vector{\phi}^{\wedge}
 \Vector{\phi}^{\wedge}\iota +
2\<\Vector{\phi}^{\wedge}
 \nuw
 \Vector{\phi}^{\wedge}
 \Vector{\phi}^{\wedge}
 \Vector{\phi}^{\wedge}
 \Vector{\phi}^{\wedge}\iota +
 \nuw
 \Vector{\phi}^{\wedge}
 \Vector{\phi}^{\wedge}
 \Vector{\phi}^{\wedge}
 \Vector{\phi}^{\wedge}
 \Vector{\phi}^{\wedge}\iota
\right) + \cdots
\end{multlined} \\[1mm]
& =
\begin{multlined}[t]
\frac{1}{3!}\<
\nuw\iota +
\frac{\phi}{4!}
\left(
2\uw
 \nuw\iota +
 \nuw
 \uw\iota
\right) +
\frac{\phi^{2}}{5!}
\left(
3\<\uw
 \uw
 \nuw\iota +
2\<\uw
 \nuw
 \uw\iota +
 \nuw
 \uw
 \uw\iota
\right) \\
+
\frac{\phi^{3}}{6!}
\left(
4\<\uw
 \uw
 \uw
 \nuw\iota +
3\<\uw
 \uw
 \nuw
 \uw\iota +
2\<\uw
 \nuw
 \uw
 \uw\iota +
 \nuw
 \uw
 \uw
 \uw\iota
\right) \\
+
\frac{\phi^{4}}{7!}
\left(
5\<\uw
 \uw
 \uw
 \uw
 \nuw\iota +
4\<\uw
 \uw
 \uw
 \nuw
 \uw\iota +
3\<\uw
 \uw
 \nuw
 \uw
 \uw\iota +
2\<\uw
 \nuw
 \uw
 \uw
 \uw\iota +
 \nuw
 \uw
 \uw
 \uw
 \uw\iota
\right) \\
+
\frac{\phi^{5}}{8!}
\left(
6\<\uw
 \uw
 \uw
 \uw
 \uw
 \nuw\iota +
5\<\uw
 \uw
 \uw
 \uw
 \nuw
 \uw\iota +
4\<\uw
 \uw
 \uw
 \nuw
 \uw
 \uw\iota
\right. \\
\left.
+~
3\<\uw
 \uw
 \nuw
 \uw
 \uw
 \uw\iota +
2\<\uw
 \nuw
 \uw
 \uw
 \uw
 \uw\iota +
 \nuw
 \uw
 \uw
 \uw
 \uw
 \uw\iota
\right) + \cdots
\end{multlined}
\end{align*}
We apply \Cref{ident:skew_products_quad} early on, obtaining
\begin{align*}
\Matrix{N}_{2}
& =
\begin{multlined}[t]
\frac{1}{3!}\<
\nuw\iota +
\frac{\phi}{4!}
\left(
2\uw
 \nuw\iota +
 \nuw
 \uw\iota
\right) +
\frac{\phi^{2}}{5!}
\left(
3\<\uw
 \uw
 \nuw\iota +
2\<\uw
 \nuw
 \uw\iota +
 \nuw
 \uw
 \uw\iota
\right) \\
+
\frac{\phi^{3}}{6!}
\left(
-4\<\uw
  \nuw\iota +
 5\<\uw
  \nuw
  \uw
  \uw\iota -
  \nuw
  \uw\iota
\right)
+
\frac{\phi^{4}}{7!}
\left(
-5\<\uw
 \uw
 \nuw\iota -
 9\<\uw
  \nuw
  \uw\iota -
  \nuw
  \uw
  \uw\iota
\right) \\
+
\frac{\phi^{5}}{8!}
\left(
6\<\uw
 \nuw\iota -
14\<\uw
 \nuw
 \uw
 \uw\iota +
 \nuw
 \uw\iota
\right)
+
\frac{\phi^{6}}{9!}
\left(
 7\<\uw
 \uw
 \nuw\iota +
20\<\uw
 \nuw
 \uw\iota +
 \nuw
 \uw
 \uw\iota
\right) + \cdots
\end{multlined}
\end{align*}
Next, we separate $\Matrix{N}_{2}$ into four submatrices, each of which can be expressed (after some tedious algebra) in closed form and summed together.
Let the matrix $\Matrix{N}_{2\mathrm{A}}$ be
\begin{align}
\Matrix{N}_{2\mathrm{A}}
& =
\sum_{n = 0}^{\infty}
\frac{n + 1}{(n + 3)!}
\bigl(\phi\<\uw\bigr)^{n} \notag \\[1mm]
& =
\frac{1}{6}\Identity_{3} +
\frac{2}{4!}\phi\<
\uw +
\frac{3}{5!}\phi^{2}
\uw
\uw +
\frac{4}{6!}\phi^{3}
\uw
\uw
\uw +
\frac{5}{7!}\phi^{4}
\uw
\uw
\uw
\uw +
\cdots \notag \\
& =
\frac{1}{6}\Identity_{3} +
\left(
\frac{2}{4!}\phi -
\frac{4}{6!}\phi^{3} +
\frac{6}{8!}\phi^{5} -
\cdots
\right)
\uw +
\left(
\frac{3}{5!}\phi^{2} -
\frac{5}{7!}\phi^{4} +
\frac{7}{9!}\phi^{6} -
\cdots
\right)
\uw
\uw \notag \\[1mm]
& =
\frac{1}{6}\Identity_{3} +
\left(
\frac{2 - \phi\sin\left(\phi\right) -
2\cos\left(\phi\right)}{\phi^{3}}
\right)
\uw +
\left(
\frac{\phi^{3} + 6\<\phi + 6\<\phi\cos\left(\phi\right) -
12\sin\left(\phi\right)}{6\<\phi^{3}}
\right)
\uw
\uw,
\end{align}
which is a function of $\Vector{\phi}$ only.
%
%
%
\noindent Let the matrix $\Matrix{N}_{2\mathrm{B}}$ be
\begin{align}
\Matrix{N}_{2\mathrm{B}}
& =
\sum_{n = 1}^{\infty}
(-1)^{n-1}
\frac{
\left( n + 1\right)
\left(2n - 1\right)
\phi^{2n}}{(2n + 3)!}
\left(
\uw
\nuw
\uw
\iota
\right) \notag \\[1mm]
& =
\left(
\frac{ 2}{5!}\phi^{2} -
\frac{ 9}{7!}\phi^{4} +
\frac{20}{9!}\phi^{6} -
\frac{35}{11!}\phi^{8} +
\cdots
\right)
\uw
\nuw
\uw
\iota \notag \\[1mm]
& =
\left(
\frac{
12\sin\left(\phi\right) -
\phi^{3} -
3\phi^{2}\sin\left(\phi\right) -
12\phi\cos\left(\phi\right)}
{6\<\phi^{3}}
\right)
\uw
\nuw
\uw
\iota.
\end{align}
%
%
%
Let the matrix $\Matrix{N}_{2\mathrm{C}}$ be
\begin{align}
\Matrix{N}_{2\mathrm{C}}
& =
\sum_{n = 1}^{\infty}
(-1)^{n-1}
\frac{
\left(2n + 3\right)
\left(n\right)
\phi^{2n + 1}}{(2n + 4)!}
\left(
\uw
\nuw
\uw
\uw
\iota
\right) \notag \\[1mm]
& =
\left(
\frac{ 5}{6!}\phi^{3} -
\frac{14}{8!}\phi^{5} +
\frac{27}{10!}\phi^{7} -
\frac{44}{12!}\phi^{9} +
\cdots
\right)
\uw
\nuw
\uw
\uw
\iota \notag \\[1mm]
& =
\left(
\frac{
4 +
\phi^{2} +
\phi^{2}\cos\left(\phi\right) -
4\phi\sin\left(\phi\right) -
4\cos\left(\phi\right)
}
{2\<\phi^{3}}
\right)
\uw
\nuw
\uw
\uw
\iota.
\end{align}

\noindent Finally, let the matrix $\Matrix{N}_{2\mathrm{D}}$ be
\begin{align}
\Matrix{N}_{2\mathrm{D}}
& =
\sum_{n = 1}^{\infty}
\frac{1}{(n + 3)!}
\bigl(\phi\<\uw\bigr)^{n} \notag \\[1mm]
& =
\frac{1}{4!}\phi\<
\uw +
\frac{1}{5!}\phi^{2}
\uw
\uw +
\frac{1}{6!}\phi^{3}
\uw
\uw
\uw +
\frac{1}{7!}\phi^{4}
\uw
\uw
\uw
\uw +
\frac{1}{8!}\phi^{5}
\uw
\uw
\uw
\uw
\uw +
\cdots \notag \\[1mm]
& =
\left(
\frac{1}{4!}\phi -
\frac{1}{6!}\phi^{3} +
\frac{1}{8!}\phi^{5} -
\cdots
\right)
\uw +
\left(
\frac{1}{5!}\phi^{2} -
\frac{1}{7!}\phi^{4} +
\frac{1}{9!}\phi^{6} -
\cdots
\right)
\uw
\uw \notag \\[1mm]
& =
\left(
\frac{\phi^{2} + 2\cos\left(\phi\right) - 2}{2\<\phi^{3}}
\right)
\uw +
\left(
\frac{\phi^{3} + 6\sin\left(\phi\right) - 6\phi}{6\<\phi^{3}}
\right)
\uw
\uw,
\end{align}
which is also a function of $\Vector{\phi}$ only.\\

\noindent The complete, closed-form solution for $\Matrix{N}_{2}$ is, at last,
\begin{equation}
\Matrix{N}_{2} =
\Matrix{N}_{2\mathrm{A}}\nuw\iota +
\Matrix{N}_{2\mathrm{B}} +
\Matrix{N}_{2\mathrm{C}} +
\nuw\iota\,\Matrix{N}_{2\mathrm{D}},
\end{equation}
or, explicitly,
\begin{align}
\Matrix{N}_{2}
& =
\begin{multlined}[t]
\frac{1}{6}
\nuw\iota +
\Biggl(
\left(
\frac{2 - \phi\sin\left(\phi\right) -
2\cos\left(\phi\right)}{\phi^{3}}
\right)
\uw +
\left(
\frac{\phi^{3} + 6\<\phi + 6\<\phi\cos\left(\phi\right) -
12\sin\left(\phi\right)}{6\<\phi^{3}}
\right)
\uw
\uw
\Biggr)
\nuw\iota \\
+
\left(
\frac{
12\sin\left(\phi\right) -
\phi^{3} -
3\phi^{2}\sin\left(\phi\right) -
12\phi\cos\left(\phi\right)}
{6\<\phi^{3}}
\right)
\uw
\nuw
\uw
\iota \hspace{0.6in} \\
\hspace{0.4in}
+
\left(
\frac{
4 +
\phi^{2} +
\phi^{2}\cos\left(\phi\right) -
4\phi\sin\left(\phi\right) -
4\cos\left(\phi\right)
}
{2\<\phi^{3}}
\right)
\uw
\uw
\nuw
\uw
\iota \\
+
\nuw\iota
\Biggl(
\left(
\frac{\phi^{2} + 2\cos\left(\phi\right) - 2}{2\<\phi^{3}}
\right)
\uw +
\left(
\frac{\phi^{3} + 6\sin\left(\phi\right) - 6\phi}{6\<\phi^{3}}
\right)
\uw
\uw
\Biggr).
\end{multlined}
\end{align}
A single application of \Cref{ident:skew_products_triple} yields the slightly simplified form
\begin{align}
\begin{multlined}[b]
\Matrix{N}_{2} =
\Biggl(
\left(
\frac{2 - \phi\sin\left(\phi\right) -
2\cos\left(\phi\right)}{\phi^{3}}
\right)
\uw +
\left(
\frac{\phi + \phi\cos\left(\phi\right) -
2\sin\left(\phi\right)}{\phi^{3}}
\right)
\uw
\uw
\Biggr)
\nuw\iota \hspace{1.0in} \\
+
\left(
\frac{
4\sin\left(\phi\right) -
 \phi^{2}\sin\left(\phi\right) -
4\phi\cos\left(\phi\right)}
{2\<\phi^{3}}
\right)
\uw
\nuw
\uw
\iota \hspace{1.5in} \\
\hspace{0.8in}
+
\left(
\frac{
4 +
\phi^{2} +
\phi^{2}\cos\left(\phi\right) -
4\phi\sin\left(\phi\right) -
4\cos\left(\phi\right)
}
{2\<\phi^{3}}
\right)
\uw
\uw
\nuw
\uw
\iota \\
+
\nuw\iota
\Biggl(
\left(
\frac{\phi^{2} + 2\cos\left(\phi\right) - 2}{2\<\phi^{3}}
\right)
\uw +
\left(
\frac{\sin\left(\phi\right) - \phi}{\phi^{3}}
\right)
\uw
\uw
\Biggr).
\end{multlined}
\end{align}


%
%
%
%
%
%
%
%
%
%
%
%
%
%

\section{The Differential of the Logarithmic Map}
\label{app:inverse_jacobian}

Evaluated at $\Matrix{F} = \Matexp{\Vector{\xi}^{\wedge}}$, the differential of the logarithmic map is given by the inverse of the left Jacobian, $\smash{\Inv{\Matrix{J}_{\ell}(\Vector{\xi})}}$.
%
%
Since $\Matrix{J}_{\ell}$ is block upper triangular, its inverse shares the same sparsity pattern.
To keep the expressions manageable, we define four $3 \times 3$ submatrices $\Matrix{P}$, $\Matrix{Q}$, $\Matrix{R}$, and $\Matrix{S}$ such that
\begin{equation}
\label{eqn:SGal3_left_jacobian_inverse_structure}
\Inv{\Matrix{J}_{\ell}(\Vector{\xi})}
=
\bbm
\Inv{\Matrix{D}} &
\Matrix{Q} &
-\Matrix{S} &
-\Matrix{P}\Vector{\nu} \\[1mm]
\Zero &
\Inv{\Matrix{D}} &
-\Matrix{R} &
\Zero \\[1mm]
\Zero &
\Zero &
\Inv{\Matrix{D}} &
\Zero \\[1mm]
\Zero &
\Zero &
\Zero &
1
\ebm.
\end{equation}
Block back-substitution gives
\begin{equation}
\label{eqn:SGal3_left_jacobian_inverse_blocks}
\Matrix{P} = \Inv{\Matrix{D}}\Matrix{E},
\quad
\Matrix{Q} = \Inv{\Matrix{D}}\Matrix{L}\Inv{\Matrix{D}}\iota,
\quad
\Matrix{R} = \Inv{\Matrix{D}}\Matrix{M}\Inv{\Matrix{D}},
\quad
\Matrix{S} = \Inv{\Matrix{D}}\left(\Matrix{N} + \Matrix{L}\Inv{\Matrix{D}}\Matrix{M}\iota\right)\Inv{\Matrix{D}}.
\end{equation}

The diagonal blocks $\Inv{\Matrix{D}}$ are already available from the logarithmic map; their closed-form expression is given in \Cref{eqn:D_inverse_closed_form}.
We begin by recovering this result using scalar power series, since the same approach will be useful for the remaining blocks.
Recall from \Cref{eqn:so3_definition} that
\begin{equation*}
\Matrix{\Phi}
=
\Vector{\phi}^{\wedge}
=
\phi\<\uw.
\end{equation*}
From the definition of $\Matrix{D}$,
\begin{equation*}
\Matrix{D}
=
\sum_{n = 0}^{\infty}
\frac{1}{(n + 1)!}
\Matrix{\Phi}^{n}.
\end{equation*}
The corresponding scalar power series is
\begin{equation*}
f(x)
\Defined
\sum_{n = 0}^{\infty}
\frac{x^{n}}{(n + 1)!}
=
\frac{e^{x} - 1}{x}.
\end{equation*}
Its reciprocal is
\begin{equation}
b(x)
\Defined
\frac{1}{f(x)}
=
\frac{x}{e^{x} - 1}
=
1
-
\frac{x}{2}
+
\frac{x^{2}}{12}
-
\frac{x^{4}}{720}
+
\frac{x^{6}}{30240}
-
\cdots.
\end{equation}
This is the generating function for the Bernoulli numbers,
\begin{equation}
\label{eqn:Bernoulli_generating_function}
b(x)
=
\sum_{n = 0}^{\infty}
\frac{B_{n}}{n!}x^{n},
\end{equation}
where $B_{n}$ denotes the $n$th Bernoulli number and we use the convention $B_{1} = -1/2$.

Using the same coefficients in the corresponding matrix series,
\begin{align*}
\Inv{\Matrix{D}}
& =
\Identity_{3}
-
\frac{\phi}{2}\uw
+
\frac{\phi^{2}}{12}\uw\uw
-
\frac{\phi^{4}}{720}\uw\uw\uw\uw
+
\frac{\phi^{6}}{30240}
\uw\uw\uw\uw\uw\uw
-
\cdots
\\[1mm]
& =
\Identity_{3}
-
\frac{\phi}{2}\uw
+
\left(
\frac{\phi^{2}}{12}
+
\frac{\phi^{4}}{720}
+
\frac{\phi^{6}}{30240}
+
\cdots
\right)
\uw\uw,
\end{align*}
where we have again used \Cref{ident:skew_negative} to reduce higher powers of $\uw$.
The remaining scalar series can be recognized from
\begin{equation*}
x\cot\left(x\right)
=
1
-
\frac{x^{2}}{3}
-
\frac{x^{4}}{45}
-
\frac{2x^{6}}{945}
-
\cdots.
\end{equation*}
Setting $x = \phi/2$ gives
\begin{equation*}
1
-
\frac{\phi}{2}\cot\left(\frac{\phi}{2}\right)
=
\frac{\phi^{2}}{12}
+
\frac{\phi^{4}}{720}
+
\frac{\phi^{6}}{30240}
+
\cdots,
\end{equation*}
and therefore
\begin{equation*}
\Inv{\Matrix{D}}
=
\Identity_{3}
-
\frac{\phi}{2}\uw
+
\Biggl(
1
-
\frac{\phi}{2}\cot\left(\frac{\phi}{2}\right)
\Biggr)
\uw\uw.
\end{equation*}

We derive the remaining blocks using the same general procedure.
For the scalar power series
\begin{equation*}
a(x)
=
\sum_{n = 0}^{\infty}
a_{n}x^{n},
\end{equation*}
we associate the matrix power series
\begin{equation*}
\sum_{n = 0}^{\infty}
a_{n}\Matrix{\Phi}^{n}.
\end{equation*}
Because powers of $\Matrix{\Phi}$ commute and combine as $\Matrix{\Phi}^{j}\Matrix{\Phi}^{k} = \Matrix{\Phi}^{j + k}$, exactly as $x^{j}x^{k} = x^{j + k}$, the product of two such matrix series is the matrix series of the product of the corresponding scalar series.
In particular, $\Inv{\Matrix{D}}\Matrix{D} = \Identity_{3}$ corresponds to $b(x)f(x) = 1$.

The matrices $\Matrix{M}$ and $\Matrix{N}$ require a small extension of this idea because they contain another matrix between the powers of $\Matrix{\Phi}$.
Given a two-variable scalar series
\begin{equation*}
w(x,y)
=
\sum_{n = 0}^{\infty}
\sum_{m = 0}^{\infty}
w_{nm}\<x^{n}y^{m},
\end{equation*}
and a fixed matrix $\Matrix{Y}$, the corresponding matrix series is
\begin{equation*}
\sum_{n = 0}^{\infty}
\sum_{m = 0}^{\infty}
w_{nm}\<
\Matrix{\Phi}^{n}
\Matrix{Y}
\Matrix{\Phi}^{m}.
\end{equation*}
Left multiplication by a matrix series corresponding to $a(x)$ amounts to multiplying $w(x,y)$ by $a(x)$, while right multiplication amounts to multiplying it by $a(y)$.
This gives us a template for all of the remaining blocks.

We first consider $\Matrix{P}$.
Recall from \Cref{eqn:E_exp_map_short} that
\begin{equation*}
\Matrix{E}
=
\sum_{n = 0}^{\infty}
\frac{1}{(n + 2)!}
\Matrix{\Phi}^{n},
\end{equation*}
for which the corresponding scalar power series is
\begin{equation*}
\sum_{n = 0}^{\infty}
\frac{x^{n}}{(n + 2)!}
=
\frac{f(x) - 1}{x}.
\end{equation*}
Since $b(x)f(x) = 1$, the scalar series corresponding to $\Matrix{P} = \Inv{\Matrix{D}}\Matrix{E}$ is
\begin{equation*}
b(x)\,\frac{f(x) - 1}{x}
=
\frac{1 - b(x)}{x}
=
-\sum_{n = 1}^{\infty}
\frac{B_{n}}{n!}x^{n - 1}
=
\frac{1}{2}
-
\frac{x}{12}
+
\frac{x^{3}}{720}
-
\frac{x^{5}}{30240}
+
\cdots.
\end{equation*}
Using the same coefficients in the corresponding matrix series and reducing higher powers of $\uw$ with \Cref{ident:skew_negative} gives
\begin{align*}
\Matrix{P}
&=
\frac{1}{2}\Identity_{3}
-
\frac{\phi}{12}\uw
+
\frac{\phi^{3}}{720}\uw\uw\uw
-
\frac{\phi^{5}}{30240}\uw\uw\uw\uw\uw
+
\cdots
\\[1mm]
&=
\frac{1}{2}\Identity_{3}
-
\left(
\frac{\phi}{12}
+
\frac{\phi^{3}}{720}
+
\frac{\phi^{5}}{30240}
+
\cdots
\right)
\uw.
\end{align*}
The remaining scalar series is the series for $1 - \frac{\phi}{2}\cot\left(\frac{\phi}{2}\right)$ from above, divided by $\phi$, and so
\begin{equation*}
\Matrix{P}
=
\frac{1}{2}\Identity_{3}
-
\Biggl(
\frac{1}{\phi}
-
\frac{1}{2}\cot\left(\frac{\phi}{2}\right)
\Biggr)
\uw.
\end{equation*}

Next, we consider $\Matrix{Q}$.
Recall from \Cref{app:jacobian} that
\begin{equation*}
\Matrix{L}
=
\sum_{n = 0}^{\infty}
\frac{n + 1}{(n + 2)!}
\Matrix{\Phi}^{n},
\end{equation*}
for which the corresponding scalar power series is
\begin{equation*}
\sum_{n = 0}^{\infty}
\frac{n + 1}{(n + 2)!}x^{n}
=
\frac{df(x)}{dx},
\end{equation*}
the derivative of $f(x)$.
Since $b(x) = 1/f(x)$, we have $\frac{db(x)}{dx} = -\frac{1}{f(x)^{2}}\frac{df(x)}{dx}$, and so the scalar series corresponding to $\Inv{\Matrix{D}}\Matrix{L}\Inv{\Matrix{D}}$ is
\begin{equation}
\label{eqn:L_scalar_relation}
b(x)\frac{df(x)}{dx}b(x)
=
-\frac{db(x)}{dx}
=
-\sum_{n = 1}^{\infty}
\frac{B_{n}}{(n - 1)!}x^{n - 1}
=
\frac{1}{2}
-
\frac{x}{6}
+
\frac{x^{3}}{180}
-
\frac{x^{5}}{5040}
+
\cdots.
\end{equation}
Using the same coefficients in the corresponding matrix series and reducing higher powers of $\uw$ with \Cref{ident:skew_negative} gives
\begin{align*}
\Matrix{Q}
&=
\Biggl(
\frac{1}{2}\Identity_{3}
-
\frac{\phi}{6}\uw
+
\frac{\phi^{3}}{180}\uw\uw\uw
-
\frac{\phi^{5}}{5040}\uw\uw\uw\uw\uw
+
\cdots
\Biggr)\iota
\\[1mm]
&=
\Biggl(
\frac{1}{2}\Identity_{3}
-
\left(
\frac{\phi}{6}
+
\frac{\phi^{3}}{180}
+
\frac{\phi^{5}}{5040}
+
\cdots
\right)
\uw
\Biggr)\iota.
\end{align*}
The remaining scalar series is the derivative, with respect to $\phi$, of the series for $1 - \frac{\phi}{2}\cot\left(\frac{\phi}{2}\right)$ from above, so that
\begin{equation*}
\frac{\phi}{6}
+
\frac{\phi^{3}}{180}
+
\frac{\phi^{5}}{5040}
+
\cdots
=
\frac{\phi}{4}\csc^{2}\left(\frac{\phi}{2}\right)
-
\frac{1}{2}\cot\left(\frac{\phi}{2}\right)
=
\frac{\phi - \sin\left(\phi\right)}
{2\left(1 - \cos\left(\phi\right)\right)},
\end{equation*}
and therefore
\begin{equation*}
\Matrix{Q}
=
\left(
\frac{1}{2}\Identity_{3}
-
\frac{\phi - \sin\left(\phi\right)}
{2\left(1 - \cos\left(\phi\right)\right)}
\uw
\right)\iota.
\end{equation*}

We now turn to $\Matrix{R}$, which requires the two-variable template.
Recall from \Cref{app:jacobian} that
\begin{equation*}
\Matrix{M}
=
\sum_{n = 0}^{\infty}
\sum_{m = 0}^{\infty}
\frac{1}{(n + m + 2)!}
\Matrix{\Phi}^{n}
\nuw
\Matrix{\Phi}^{m},
\end{equation*}
for which the corresponding two-variable scalar series is
\begin{equation*}
h(x,y)
\Defined
\sum_{n = 0}^{\infty}
\sum_{m = 0}^{\infty}
\frac{x^{n}y^{m}}{(n + m + 2)!}
=
\frac{f(x) - f(y)}{x - y}.
\end{equation*}
When $x = y$, the right-hand side takes its limiting value, $\frac{df(x)}{dx}$.
The scalar series corresponding to $\Matrix{R} = \Inv{\Matrix{D}}\Matrix{M}\Inv{\Matrix{D}}$ is then, using \Cref{eqn:Bernoulli_generating_function},
\begin{align}
c(x,y)
& \Defined
b(x)h(x,y)b(y)
=
\frac{b(y) - b(x)}{x - y}
=
-\sum_{n = 1}^{\infty}
\frac{B_{n}}{n!}
\sum_{m = 0}^{n - 1}
x^{n - 1 - m}y^{m}
\notag \\[1mm]
& =
\frac{1}{2}
-
\frac{x + y}{12}
+
\frac{x^{3} + x^{2}y + xy^{2} + y^{3}}{720}
-
\frac{x^{5} + x^{4}y + x^{3}y^{2} + x^{2}y^{3} + xy^{4} + y^{5}}{30240}
+
\cdots.
\label{eqn:R_scalar_series}
\end{align}
Using the same coefficients in the corresponding matrix series gives
\begin{equation*}
\Matrix{R}
=
\frac{1}{2}\nuw
-
\frac{1}{12}
\left(
\Matrix{\Phi}\nuw
+
\nuw\Matrix{\Phi}
\right)
+
\frac{1}{720}
\left(
\Matrix{\Phi}^{3}\nuw
+
\Matrix{\Phi}^{2}\nuw\Matrix{\Phi}
+
\Matrix{\Phi}\nuw\Matrix{\Phi}^{2}
+
\nuw\Matrix{\Phi}^{3}
\right)
-
\cdots.
\end{equation*}
Substituting $\Matrix{\Phi} = \phi\<\uw$ and reducing with \Cref{ident:skew_negative,ident:skew_products_quad} leaves two scalar series,
\begin{equation*}
\Matrix{R}
=
\frac{1}{2}\nuw
-
\left(
\frac{\phi}{12}
+
\frac{\phi^{3}}{720}
+
\frac{\phi^{5}}{30240}
+
\cdots
\right)
\left(
\uw\nuw
+
\nuw\uw
\right)
+
\left(
\frac{\phi^{3}}{360}
+
\frac{\phi^{5}}{7560}
+
\cdots
\right)
\uw\nuw\uw\uw.
\end{equation*}
The first is the series for $\Matrix{P}$, and the second is the series for $\Matrix{Q}$ minus twice the series for $\Matrix{P}$, so that
\begin{equation*}
\frac{\phi^{3}}{360}
+
\frac{\phi^{5}}{7560}
+
\cdots
=
\Biggl(
\frac{\phi}{4}\csc^{2}\left(\frac{\phi}{2}\right)
-
\frac{1}{2}\cot\left(\frac{\phi}{2}\right)
\Biggr)
-
2\Biggl(
\frac{1}{\phi}
-
\frac{1}{2}\cot\left(\frac{\phi}{2}\right)
\Biggr)
=
-
\Biggl(
\frac{2}{\phi}
-
\frac{1}{2}\cot\left(\frac{\phi}{2}\right)
-
\frac{\phi}{4}\csc^{2}\left(\frac{\phi}{2}\right)
\Biggr),
\end{equation*}
and therefore
\begin{equation*}
\Matrix{R}
=
\frac{1}{2}\nuw
-
\Biggl(
\frac{1}{\phi}
-
\frac{1}{2}\cot\left(\frac{\phi}{2}\right)
\Biggr)
\left(
\uw\nuw
+
\nuw\uw
\right)
-
\Biggl(
\frac{2}{\phi}
-
\frac{1}{2}\cot\left(\frac{\phi}{2}\right)
-
\frac{\phi}{4}\csc^{2}\left(\frac{\phi}{2}\right)
\Biggr)
\uw\nuw\uw\uw.
\end{equation*}

Finally, we consider $\Matrix{S}$.
Recall from \Cref{app:jacobian} that $\Matrix{N} = \Matrix{N}_{1} - \Matrix{N}_{2}$, where
\begin{equation*}
\Matrix{N}_{1}
=
\sum_{n = 0}^{\infty}
\sum_{m = 0}^{\infty}
\frac{1}{(n + m + 2)!}
\Matrix{\Phi}^{n}
\puw
\Matrix{\Phi}^{m}
\quad \text{and} \quad
\Matrix{N}_{2}
=
\sum_{n = 0}^{\infty}
\sum_{m = 0}^{\infty}
\frac{n + 1}{(n + m + 3)!}
\Matrix{\Phi}^{n}
\nuw
\Matrix{\Phi}^{m}
\iota.
\end{equation*}
The part of $\Matrix{S}$ involving $\Vector{\rho}$ is $\Inv{\Matrix{D}}\Matrix{N}_{1}\Inv{\Matrix{D}}$, whose scalar series is again $c(x,y)$ from \Cref{eqn:R_scalar_series}; it therefore has the same form as $\Matrix{R}$, with $\nuw$ replaced by $\puw$.
The part involving $\Vector{\nu}$ is $-\Inv{\Matrix{D}}\Matrix{N}_{2}\Inv{\Matrix{D}} + \Inv{\Matrix{D}}\Matrix{L}\Inv{\Matrix{D}}\Matrix{M}\Inv{\Matrix{D}}\iota$.
The scalar series for $\Matrix{N}_{2}$ is $\partial h(x,y)/\partial x$, and that for $\Inv{\Matrix{D}}\Matrix{L}\Inv{\Matrix{D}}$ is $-db(x)/dx$ from \Cref{eqn:L_scalar_relation}, so the scalar series for this part is
\begin{equation*}
-b(x)\frac{\partial h(x,y)}{\partial x}b(y)
-
\frac{db(x)}{dx}h(x,y)b(y)
=
-\frac{\partial c(x,y)}{\partial x},
\end{equation*}
by the product rule applied to $c(x,y) = b(x)h(x,y)b(y)$.
Differentiating \Cref{eqn:R_scalar_series} gives
\begin{equation*}
-\frac{\partial c(x,y)}{\partial x}
=
\frac{1}{12}
-
\frac{3x^{2} + 2xy + y^{2}}{720}
+
\frac{5x^{4} + 4x^{3}y + 3x^{2}y^{2} + 2xy^{3} + y^{4}}{30240}
-
\cdots.
\end{equation*}
Using the same coefficients in the corresponding matrix series, substituting $\Matrix{\Phi} = \phi\<\uw$, and reducing with \Cref{ident:skew_negative,ident:skew_products_triple,ident:skew_products_quad} leaves three scalar series in the part involving $\Vector{\nu}$,
\begin{equation*}
\left(
\frac{1}{12}
+
\frac{\phi^{2}}{720}
+
\frac{\phi^{4}}{30240}
+
\cdots
\right)
\nuw\iota
-
\left(
\frac{\phi^{2}}{360}
+
\frac{\phi^{4}}{7560}
+
\cdots
\right)
\uw\uw\nuw\iota
-
\left(
\frac{\phi^{2}}{240}
+
\frac{\phi^{4}}{3024}
+
\cdots
\right)
\uw\nuw\uw\iota.
\end{equation*}
The first two are the two series from $\Matrix{R}$, each divided by $\phi$, and the third is
\begin{equation*}
\frac{\phi^{2}}{240}
+
\frac{\phi^{4}}{3024}
+
\cdots
=
\frac{1}{\phi^{2}}
-
\frac{\phi}{8}\cot\left(\frac{\phi}{2}\right)\csc^{2}\left(\frac{\phi}{2}\right),
\end{equation*}
which can be checked by expanding the right-hand side.
Combining the two parts gives
\begin{align*}
\Matrix{S}
&
\begin{multlined}[t]
=
\frac{1}{2}\puw
-
\Biggl(
\frac{1}{\phi}
-
\frac{1}{2}\cot\left(\frac{\phi}{2}\right)
\Biggr)
\left(
\uw\puw
+
\puw\uw
\right)
\\[1mm]
\shoveleft[0.25in]{
-
\Biggl(
\frac{2}{\phi}
-
\frac{1}{2}\cot\left(\frac{\phi}{2}\right)
-
\frac{\phi}{4}\csc^{2}\left(\frac{\phi}{2}\right)
\Biggr)
\uw\puw\uw\uw
+
\frac{1}{\phi}
\Biggl(
\frac{1}{\phi}
-
\frac{1}{2}\cot\left(\frac{\phi}{2}\right)
\Biggr)
\nuw\iota
}
\\[1mm]
\hspace{0.5in}
+
\frac{1}{\phi}
\Biggl(
\frac{2}{\phi}
-
\frac{1}{2}\cot\left(\frac{\phi}{2}\right)
-
\frac{\phi}{4}\csc^{2}\left(\frac{\phi}{2}\right)
\Biggr)
\uw\uw\nuw\iota
+
\Biggl(
\frac{\phi}{8}\cot\left(\frac{\phi}{2}\right)\csc^{2}\left(\frac{\phi}{2}\right)
-
\frac{1}{\phi^{2}}
\Biggr)
\uw\nuw\uw\iota.
\end{multlined}
\end{align*}

At $\phi = 0$, the closed-form expressions above take their limiting values,
\begin{equation*}
\Inv{\Matrix{D}} = \Identity_{3},
\quad
\Matrix{P} = \frac{1}{2}\Identity_{3},
\quad
\Matrix{Q} = \frac{1}{2}\Identity_{3}\iota,
\quad
\Matrix{R} = \frac{1}{2}\nuw,
\quad
\Matrix{S} = \frac{1}{2}\puw + \frac{1}{12}\nuw\iota.
\end{equation*}

\section{Some Useful Identities}
\label{app:identities}

This appendix contains several useful identities for manipulating expressions involving products of skew-symmetric matrices.

\vspace{0.25\baselineskip}
\begin{identity}
\label{ident:skew_negative}
Let $\Vector{u} \in \Real^{3}$ and let $\Norm{\Vector{u}} = 1$. Then the following identity holds:
\begin{equation}
\label{eqn:identity_1}
\uw\uw\uw = -\uw.
\end{equation}
\end{identity}

\begin{proof}
Let $\Vector{u} = \bbm u_{1}\nms & u_{2}\nms & u_{3} \ebm^{\T}$. The matrix $\uw\uw$ is symmetric and
\begin{align*}
\uw\uw\uw
& =
\underbrace{%
\bbm
u_{1}^{2} - 1 &    u_{1}u_{2} &    u_{1}u_{3} \\
   u_{1}u_{2} & u_{2}^{2} - 1 &    u_{2}u_{3} \\
   u_{1}u_{3} &    u_{2}u_{3} & u_{3}^{2} - 1
\ebm
}_{\uw\uw}
\underbrace{%
\bbm
     0 & -u_{3} &  u_{2} \\
 u_{3} &      0 & -u_{1} \\
-u_{2} &  u_{1} &      0
\ebm
}_{\uw} \\
& = 
\bbm
u_{1}u_{2}u_{3} - u_{1}u_{2}u_{3} &
u_{1}^{2}u_{3} - u_{1}^{2}u_{3} + u_{3} &
u_{1}^{2}u_{2} - u_{1}^{2}u_{2} - u_{2} \\
u_{2}^{2}u_{3} - u_{2}^{2}u_{3} - u_{3} &
u_{1}u_{2}u_{3} - u_{1}u_{2}u_{3} & 
u_{1}u_{2}^{2} - u_{1}u_{2}^{2} + u_{1} \\
u_{2}u_{3}^{2} - u_{2}u_{3}^{2} + u_{2} &
u_{1}u_{3}^{2} - u_{1}u_{3}^{2} - u_{1} &
u_{1}u_{2}u_{3} - u_{1}u_{2}u_{3}
\ebm
=
\bbm
     0 &  u_{3} & -u_{2} \\
-u_{3} &      0 &  u_{1} \\
 u_{2} & -u_{1} &      0
\ebm
=
-\uw.
\end{align*}
\end{proof}

\vspace{0.25\baselineskip}
\begin{identity}
\label{ident:skew_products_triple}
Let $\Vector{u}, \Vector{v} \in \Real^{3}$ and let $\Norm{\Vector{u}} = 1$. Then the following identity holds:
\begin{equation}
\label{eqn:identity_2}
\uw\Vector{v}^{\wedge}\uw - 
\uw\uw\Vector{v}^{\wedge} -
\Vector{v}^{\wedge}\uw\uw 
= 
\Vector{v}^{\wedge}.
\end{equation}
\end{identity}

\begin{proof}
Let $\Vector{u} = \bbm u_{1}\nms & u_{2}\nms & u_{3} \ebm^{\T}$ and  $\Vector{v} = \bbm v_{1}\nms & v_{2}\nms & v_{3} \ebm^{\T}$. We have
\begin{align*}
\uw\Vector{v}^{\wedge}\uw
 & = 
\bbm
 0 & 
 \left(u_{1}v_{1} + u_{2}v_{2} + u_{3}v_{3}\right)u_{3} &
-\left(u_{1}v_{1} + u_{2}v_{2} + u_{3}v_{3}\right)u_{2} \\
-\left(u_{1}v_{1} + u_{2}v_{2} + u_{3}v_{3}\right)u_{3} &
 0 &
 \left(u_{1}v_{1} + u_{2}v_{2} + u_{3}v_{3}\right)u_{1} \\
 \left(u_{1}v_{1} + u_{2}v_{2} + u_{3}v_{3}\right)u_{2} &
-\left(u_{1}v_{1} + u_{2}v_{2} + u_{3}v_{3}\right)u_{1} &
 0
\ebm, \\[3mm]
\uw\uw\Vector{v}^{\wedge}
& = 
\bbm
-u_{1}u_{3}v_{2} + u_{1}u_{2}v_{3} &
 u_{1}u_{3}v_{1} - (u_{1}^{2} - 1)v_{3} &
-u_{1}u_{2}v_{1} + (u_{1}^{2} - 1)v_{2} \\
-u_{2}u_{3}v_{2} + (u_{2}^{2} - 1)v_{3} & 
 u_{2}u_{3}v_{1} - u_{1}u_{2}v_{3}  & 
 u_{1}u_{2}v_{2} - (u_{2}^{2} - 1)v_{1} \\
 u_{2}u_{3}v_{3} - (u_{3}^{2} - 1)v_{2} &
-u_{1}u_{3}v_{3} + (u_{3}^{2} - 1)v_{1} &
-u_{2}u_{3}v_{1} + u_{1}u_{3}v_{2}
\ebm, \\[3mm]
\Vector{v}^{\wedge}\uw\uw
& =
\bbm
 u_{1}u_{3}v_{2} - u_{1}u_{2}v_{3} &
 (u_{2}v_{2} + u_{3}v_{3})u_{3} + u_{1}^{2}v_{3} &
-(u_{2}v_{2} + u_{3}v_{3})u_{2} - u_{1}^{2}v_{2} \\
-(u_{1}v_{1} + u_{3}v_{3})u_{3} - u_{2}^{2}v_{3} &
-u_{2}u_{3}v_{1} + u_{1}u_{2}v_{3} &
 (u_{1}v_{1} + u_{3}v_{3})u_{1} + u_{2}^{2}v_{1} \\
 (u_{1}v_{1} + u_{2}v_{2})u_{2} + u_{3}^{2}v_{2} &
-(u_{1}v_{1} + u_{2}v_{2})u_{1} - u_{3}^{2}v_{1} &
 u_{2}u_{3} v_{1} - u_{1}u_{3}v_{2}
\ebm.
\end{align*}
The result follows directly.
\end{proof}

\vspace{0.25\baselineskip}
\begin{identity}
\label{ident:skew_products_quad}
Let $\Vector{u}, \Vector{v} \in \Real^{3}$ and let $\Norm{\Vector{u}} = 1$. Then the following identity holds:
\begin{equation}
\label{eqn:identity_3}
\uw\uw\Vector{v}^{\wedge}\uw
=
\uw\Vector{v}^{\wedge}\uw\uw.
\end{equation}
\end{identity}

\begin{proof}
Let $\Vector{u} = \bbm u_{1}\nms & u_{2}\nms & u_{3} \ebm^{\T}$ and  $\Vector{v} = \bbm v_{1}\nms & v_{2}\nms & v_{3} \ebm^{\T}$. We have
\begin{multline*}
\uw\uw\Vector{v}^{\wedge}\uw
=
\underbrace{%
\bbm
u_{1}^{2} - 1 &    u_{1}u_{2} &    u_{1}u_{3} \\
   u_{1}u_{2} & u_{2}^{2} - 1 &    u_{2}u_{3} \\
   u_{1}u_{3} &    u_{2}u_{3} & u_{3}^{2} - 1
\ebm
}_{\uw\uw}
\underbrace{%
\bbm
-u_{2}v_{2} - u_{3}v_{3} &
 u_{1}v_{2} &
 u_{1}v_{3} \\
 u_{2}v_{1} &
-u_{1}v_{1} - u_{3}v_{3} &
 u_{2}v_{3} \\
 u_{3}v_{1} &
 u_{3}v_{2} &
-u_{1}v_{1} - u_{2}v_{2}
\ebm
}_{\Vector{v}^{\wedge}\uw} \\[1mm]
= 
\bbm
 \left(u_{1}v_{1} + u_{2}v_{2} + u_{3}v_{3}\right)
 \left(1 - u_{1}^{2}\right) &
-\left(u_{1}v_{1} + u_{2}v_{2} + u_{3}v_{3}\right)u_{1}u_{2} &
-\left(u_{1}v_{1} + u_{2}v_{2} + u_{3}v_{3}\right)u_{1}u_{3} \\
-\left(u_{1}v_{1} + u_{2}v_{2} + u_{3}v_{3}\right)u_{1}u_{2} &
 \left(u_{1}v_{1} + u_{2}v_{2} + u_{3}v_{3}\right)
 \left(1 - u_{2}^{2}\right) &
-\left(u_{1}v_{1} + u_{2}v_{2} + u_{3}v_{3}\right)u_{2}u_{3} \\
-\left(u_{1}v_{1} + u_{2}v_{2} + u_{3}v_{3}\right)u_{1}u_{3} &
-\left(u_{1}v_{1} + u_{2}v_{2} + u_{3}v_{3}\right)u_{2}u_{3} & 
 \left(u_{1}v_{1} + u_{2}v_{2} + u_{3}v_{3}\right)
 \left(1 - u_{3}^{2}\right)
\ebm
\end{multline*}
and
\begin{multline*}
\uw\Vector{v}^{\wedge}\uw\uw
=
\underbrace{%
\bbm
-u_{2}v_{2} - u_{3}v_{3} &
 u_{2}v_{1} &
 u_{3}v_{1} \\
 u_{1}v_{2} &
-u_{1}v_{1} - u_{3}v_{3} &
 u_{3}v_{2} \\
 u_{1}v_{3} &
 u_{2}v_{3} &
-u_{1}v_{1} - u_{2}v_{2}
\ebm
}_{\uw\Vector{v}^{\wedge}}
\underbrace{%
\bbm
u_{1}^{2} - 1 &    u_{1}u_{2} &    u_{1}u_{3} \\
   u_{1}u_{2} & u_{2}^{2} - 1 &    u_{2}u_{3} \\
   u_{1}u_{3} &    u_{2}u_{3} & u_{3}^{2} - 1
\ebm
}_{\uw\uw} \\[1mm]
= 
\bbm
 \left(u_{1}v_{1} + u_{2}v_{2} + u_{3}v_{3}\right)
 \left(1 - u_{1}^{2}\right) &
-\left(u_{1}v_{1} + u_{2}v_{2} + u_{3}v_{3}\right)u_{1}u_{2} &
-\left(u_{1}v_{1} + u_{2}v_{2} + u_{3}v_{3}\right)u_{1}u_{3} \\
-\left(u_{1}v_{1} + u_{2}v_{2} + u_{3}v_{3}\right)u_{1}u_{2} &
 \left(u_{1}v_{1} + u_{2}v_{2} + u_{3}v_{3}\right)
 \left(1 - u_{2}^{2}\right) &
-\left(u_{1}v_{1} + u_{2}v_{2} + u_{3}v_{3}\right)u_{2}u_{3} \\
-\left(u_{1}v_{1} + u_{2}v_{2} + u_{3}v_{3}\right)u_{1}u_{3} &
-\left(u_{1}v_{1} + u_{2}v_{2} + u_{3}v_{3}\right)u_{2}u_{3} & 
 \left(u_{1}v_{1} + u_{2}v_{2} + u_{3}v_{3}\right)
 \left(1 - u_{3}^{2}\right)
\ebm.
\end{multline*}

\vspace{0.25\baselineskip}
\noindent Alternatively, \Cref{ident:skew_products_quad} can be proved by making use of \Cref{ident:skew_products_triple} and \Cref{ident:skew_negative}. Rearranging \Cref{eqn:identity_2} and right-multiplying by $\uw$ gives
\begin{align*}
\uw\uw\Vector{v}^{\wedge}\uw 
& =
\left(
\uw\Vector{v}^{\wedge}\uw - 
\Vector{v}^{\wedge}\uw\uw -
\Vector{v}^{\wedge}
\right)
\uw \\[1mm]
& =
\uw\Vector{v}^{\wedge}\uw\uw - 
\Vector{v}^{\wedge}\uw\uw\uw -
\Vector{v}^{\wedge}\uw \\
& =
\uw\Vector{v}^{\wedge}\uw\uw + 
\Vector{v}^{\wedge}\uw -
\Vector{v}^{\wedge}\uw \\
& = 
\uw\Vector{v}^{\wedge}\uw\uw.
\end{align*}
\end{proof}

\section{Revision History}
\label{app:revisions}

A rough summary of revisions to the report follows. Only major changes and corrections are listed.

\begin{itemize}
\setlength{\itemsep}{1.5pt}
\item Revision 1.10, 2023-11-26 --- Initial release.
\item Revision 1.11, 2023-12-12 --- Fixed error in Lie group adjoint.
\item Revision 1.12, 2024-01-07 --- Added material on Lie bracket and Lie algebra adjoint.
\item Revision 1.14, 2024-01-17 --- Added material on left Jacobian.
\item Revision 1.17, 2024-02-06 --- Fixed two errors in left Jacobian expression.
\item Revision 1.20, 2024-02-25 --- Fixed missing terms in closed-form expression for left Jacobian.
\item Revision 1.21, 2024-03-17 --- Added several references and literature pointers.
\item Revision 1.22, 2024-03-29 --- Added brief notes on simultaneity in Galilean spacetime.
\item Revision 1.23, 2024-08-18 --- Fixed missing logarithm typo in the covariance expression in \Cref{eqn:sgal3_covariance}.
\item Revision 1.24, 2024-08-24 --- Removed appendix with power series (commented out).
\item Revision 1.25, 2024-08-25 --- Made minor edits to \Cref{sec:preliminaries} regarding Lie bracket.
\item Revision 1.26, 2024-12-04 --- Added \Cref{sec:galilean_relativity} with more material on Galilean relativity.
\item Revision 1.27, 2025-01-05 --- Added reference to L\'{e}vy-Leblond (1971).
\item Revision 1.28, 2025-02-05 --- Added \Cref{app:identities}, containing identities useful for computing Jacobians.
\item Revision 1.30, 2025-02-23 --- Used \Cref{app:identities} to simplify $\LieGroupSE{3}$ and $\LieGroupSGal{3}$ Jacobians.
\item Revision 1.31, 2025-05-30 --- Edits to some definitions and surrounding text for concision and clarity.
\item Revision 1.32, 2025-06-05 --- Further simplified $\LieGroupSGal{3}$ Jacobian.
\item Revision 1.33, 2025-06-17 --- Tidied up some internals.
\item Revision 1.34, 2025-06-26 --- Minor clarifications and formatting adjustments.
\item Revision 1.35, 2026-02-12 --- Small typographical corrections throughout.
\item Revision 1.36, 2026-04-30 --- Removed event transform example.
\item Revision 1.37, 2026-06-01 --- Added differential of logarithmic map in \Cref{sec:logmap_differential} and \Cref{app:inverse_jacobian}.
\item Revision 1.38, 2026-06-10 --- Revised material on Galilean spacetime and inertial frames.
\item Revision 1.39, 2026-09-21 --- Fixed missing adjoint operator in Jacobian in \Cref{sec:adjoint}.
\item Revision 1.40, 2026-10-10 --- Added the Cayley and inverse Cayley maps in \Cref{sec:cayley_maps}.
\item Revision 1.41, 2026-10-13 --- Small typographical corrections throughout.
\item Revision 1.42, 2026-10-24 --- Tidied up some internals again.
\item Revision 1.43, 2026-10-28 --- Simplified inverse left Jacobian and rewrote its derivation in \Cref{app:inverse_jacobian}.
\end{itemize}

\printbibliography

\end{document}